\documentclass[aps,pre,twocolumn,10pt,showpacs,superscriptaddress,nofootinbib]{revtex4-1}
\usepackage{amsmath,amssymb}
\usepackage{graphicx,color,colortbl}
\usepackage{rotating,array,tabularx,booktabs}
\newcolumntype{Y}{>{\centering\arraybackslash}X}

\newcommand{\dif}{\mathrm{d}}%
\newcommand{\fdif}{\operatorname{\delta}}
\newcommand{\Fdif}[2]{\frac{\fdif\!#1}{\fdif\!#2}}
\newcommand{\ii}{\mathrm{i}}%
\newcommand{\Nabla}{\vec{\nabla}}%
\newcommand{\R}{\mathbb{R}}%
\newcommand{\Tr}{\operatorname{Tr}}%
\newcommand{\ie}{i.\,e.}%
\newcommand{\ZT}[1]{\textquotedblleft#1\textquotedblright}%
\newcommand{\rs}{\vec{r}\hskip1pt'}%
\newcommand{\rss}{\vec{r}\hskip1pt''}%
\newcommand{\HO}{\hat{H}}%
\newcommand{\Heff}{\hat{H}_{\mathrm{eff}}}%

\begin{document}
\title{Microscopic approach to entropy production}

\author{Raphael Wittkowski}\author{Hartmut L{\"o}wen}
\affiliation{Institut f{\"u}r Theoretische Physik II, Weiche Materie,
Heinrich-Heine-Universit{\"a}t D{\"u}sseldorf, D-40225 D{\"u}sseldorf, Germany}

\author{Helmut R. Brand}
\affiliation{Theoretische Physik III, Universit{\"a}t Bayreuth, D-95540 Bayreuth, Germany}

\date{\today}

\begin{abstract}
It is a great challenge of nonequilibrium statistical mechanics to calculate entropy production
within a microscopic theory.
In the framework of linear irreversible thermodynamics, we combine the Mori-Zwanzig-Forster projection operator technique 
with the first and second law of thermodynamics to obtain microscopic expressions for the entropy production as well as for the
transport equations of the entropy density and its time correlation function.
We further present  a microscopic derivation of a dissipation functional from which the dissipative dynamics 
of an extended dynamical density functional theory can be obtained in a formally elegant way.      
\end{abstract}


\maketitle


\section{\label{sec:introduction}Introduction}
One of the central challenges of statistical mechanics is to calculate and predict the entropy of a given system 
under well-specified conditions from a microscopic point of view. In equilibrium, this problem dates back to Boltzmann
and it is by now standard textbook knowledge that the entropy can be obtained by a suitable phase-space averaging 
in different ensembles \cite{Reichl1998}. The same problem, however, is much more complicated and in general unsolved in a 
nonequilibrium situation: while the second law of thermodynamics predicts a global entropy production in a closed system,
it is very difficult -- even in principle -- to calculate the entropy production within a microscopic approach that starts from the individual 
interactions of the particles. Linear irreversible thermodynamics \cite{MartinPP1972,deGrootM1984,Reichl1998}
provides a framework, where this problem can be addressed systematically in a simpler way. 
Moreover, the Mori-Zwanzig-Forster projection operator technique (MZFT) 
\cite{Mori1965,ZwanzigM1965,Forster1974,Grabert1982,Forster1990,Dhont1996,Zwanzig2001}, when supplemented with a selection of relevant slow variables,
can be used to derive microscopic expressions for the dynamical evolution of these slow variables and their time correlation functions. 
When the MZFT is applied to the entropy density, there is, however, a principle obstacle since the entropy density does not
possess any known microscopic operator. This fact is opposed to, for example, the particle density and the energy density, which both possess 
corresponding microscopic operators.

In fact, we have recently applied the MZFT in the framework of linear irreversible thermodynamics to derive a generalized (extended)
dynamical density functional theory (EDDFT), which provides a microscopic basis for the dynamical evolution equations of the relevant variables 
\cite{WittkowskiLB2012}. The approach was only formulated for slow conserved variables such as particle and energy density.
In this paper, we generalize the EDDFT approach of Ref.\ \cite{WittkowskiLB2012} to include the entropy density $\sigma(\vec{r},t)$ 
as an additional (non-conserved) variable.
Assuming the applicability of linear irreversible thermodynamics \cite{MartinPP1972,deGrootM1984,Reichl1998}, 
we derive transport equations for the entropy density as well as for its time correlation function $C_{\sigma\sigma}(\vec{r},\rs\!,t,t')$.
The key idea here is to combine the first and second law of thermodynamics with the microscopic expressions obtained
within the MZFT for the remaining slow variables in order to circumvent the obstacle that the entropy density does not possess 
a known microscopic operator. 
Thereby, we provide -- at least in principle -- a link of entropy production to microscopic expressions. 
In doing so, we also present a microscopic expression for a dissipation functional from which EDDFT \cite{WittkowskiLB2012} and 
in particular the traditional dynamical density functional theory (DDFT) \cite{MarconiT1999,MarconiT2000,ArcherE2004,EspanolL2009,WittkowskiL2011} 
can be derived in an elegant way. An important application of this reformulation of EDDFT in terms of a dissipation functional is the derivation of 
phase field crystal (PFC) models from EDDFT, in particular those which involve orientational degrees of freedom
\cite{WittkowskiLB2010,WittkowskiLB2011,WittkowskiLB2011b,EmmerichEtAl2012}. 

Our derivation of the transport equations for the entropy density and its time correlation function is valid for a general set of 
relevant variables including also variables that are not considered in the context of current EDDFT. 
Furthermore, the derived expressions are not restricted to the hydrodynamic limit 
(vanishing wave vector $\vec{k}\to\vec{0}$ and frequency $\omega\to 0$) and thus generalize the corresponding hydrodynamic equations 
to larger wave vectors and frequencies. 

The paper is organized as follows: after preliminary remarks about the considered systems and notation in Sec.\ \ref{sec:RV},
we present a derivation of transport equations for the entropy density and its time correlation function as well as a 
microscopic expression for the corresponding dissipation functional in Sec.\ \ref{sec:ED}.
These expressions for general relevant variables are afterwards specialized to the variables of EDDFT in Sec.\ \ref{sec:EDDFT}.
Finally, we conclude in Sec.\ \ref{sec:conclusions}.

\section{\label{sec:RV}Relevant variables and their transport equations}
As usual in the context of EDDFT \cite{WittkowskiLB2012}, we consider a grand-canonical (total) ensemble of systems of 
$N$ particles with the not explicitly time-dependent Hamiltonian $\HO(\hat{\Gamma}_{t})$\footnote{For consistency, we denote all quantities 
that are directly associated with the set of phase-space variables $\hat{\Gamma}_{t}$ by a hat $\hat{\,\,\,}$, 
while the hat is omitted for all other quantities.} and the set $\hat{\Gamma}_{t}$ of phase-space coordinates.
These systems can be described by the total probability density $\hat{\rho}(t)$, which is given by the solution of the 
Liouville-von Neumann equation
\begin{equation}
\dot{\hat{\rho}}=-\hat{\mathcal{L}}\:\!\hat{\rho}=-\frac{\ii}{\hbar}[\HO,\hat{\rho}] \;,\qquad \hat{\rho}(t)=e^{-\hat{\mathcal{L}}t}\hat{\rho}(0)
\label{eq:Liouville_rho}
\end{equation}
with the Liouvillian $\hat{\mathcal{L}}$, the imaginary unit $\ii$, the reduced Planck constant $\hbar=h/(2\pi)$, 
and the commutator $[X,Y]=XY-YX$ of $X$ and $Y$.

\subsection{Relevant variables}
Following the idea of the MZFT, we select a set $\Gamma_{t}$ of $n\ll N$ independent 
relevant variables (operators) $\hat{a}_{i}(\vec{r},t)$ with $i\in\{1,\dotsc,n\}$, which is sufficient to describe the considered system 
on a much simpler basis, where the huge number of irrelevant microscopic variables is projected out. 
Transport equations for the relevant variables are given by the Liouville-von Neumann equations 
\begin{equation}
\dot{\hat{a}}_{i}=\hat{\mathcal{L}}\:\!\hat{a}_{i}=\frac{\ii}{\hbar}[\HO,\hat{a}_{i}] \;,\qquad 
\hat{a}_{i}(\vec{r},t)=e^{\hat{\mathcal{L}}t}\hat{a}_{i}(\vec{r},0) \;.
\label{eq:Liouville_A}
\end{equation}
The set of relevant variables $\Gamma_{t}$ is associated with a relevant probability density $\rho(t)$. 
Together with the grand-ca\-no\-ni\-cal trace $\Tr$, the relevant probability density can be used to
define the averaged relevant variables 
\begin{equation}
a_{i}(\vec{r},t)=\Tr(\rho(t) \hat{a}_{i}(\vec{r},0)) \;.
\end{equation}
We assume a given generalized Helmholtz free-energy functional\footnote{Static stability of the considered system requires 
that $\mathcal{F}$ is bounded from below.} 
\begin{equation}
\mathcal{F}[a_{1},\dotsc,a_{n}]=\int_{\R^{3}}\!\!\!\!\!\:\!\dif^{3}r\, f(\vec{r},t)
\label{eq:F}%
\end{equation}
with the generalized Helmholtz free-energy density $f(\vec{r},t)$ that describes the state of the considered system in terms of the 
averaged relevant variables $a_{i}(\vec{r},t)$.
The generalized Helmholtz free-energy functional \eqref{eq:F} can be used to define the \textit{thermodynamic conjugates}\footnote{The 
internal energy density is an exception. Its thermodynamic conjugate has to be considered differently 
(see Secs.\ \ref{subsec:TE} and \ref{subsec:NED}).} 
\begin{equation}
a^{\natural}_{i}(\vec{r},t)=\Fdif{\mathcal{F}[a_{1},\dotsc,a_{n}]}{a_{i}(\vec{r},t)}
\end{equation}
of the averaged relevant variables $a_{i}(\vec{r},t)$.
In terms of the relevant variables and their thermodynamic conjugates, the relevant probability density $\rho(t)$ is here specified 
as the generalized grand-canonical probability density 
\begin{equation}
\rho(t)=\frac{1}{\Xi(t)}\,e^{-\beta \Heff(t)} 
\label{eq:rho}%
\end{equation}
with the grand-canonical partition sum $\Xi(t)$, 
the (constant) inverse thermal energy $\beta=1/(k_{\mathrm{B}}T)$, where $k_{\mathrm{B}}$ denotes the Boltzmann constant 
and $T$ is the absolute temperature, and the explicitly time-dependent effective Hamiltonian
\begin{equation}
\Heff(t)=\HO-\sum^{n}_{i=1}\int_{\R^{3}}\!\!\!\!\!\:\!\dif^{3}r\, a^{\natural}_{i}(\vec{r},t)\hat{a}_{i}(\vec{r}) \;.
\end{equation}
The $n=n_{\mathrm{c}}+n_{\mathrm{n}}$ relevant variables $\hat{a}_{i}(\vec{r},t)$ have to be distinguished into 
$n_{\mathrm{c}}$ conserved variables $\hat{a}^{\mathrm{c}}_{i}(\vec{r},t)$ and $n_{\mathrm{n}}$ non-conserved variables 
$\hat{a}^{\mathrm{n}}_{i}(\vec{r},t)$.
The same holds for their averages $a_{i}(\vec{r},t)$, which are distinguished into conserved averaged variables $a^{\mathrm{c}}_{i}(\vec{r},t)$ 
and non-conserved averaged variables $a^{\mathrm{n}}_{i}(\vec{r},t)$.

\subsection{\label{subsec:TE}Transport equations}
The dynamics of the $n_{\mathrm{c}}$ conserved relevant variables can be described by the conservation equations
\begin{equation}
\dot{\hat{a}}^{\mathrm{c}}_{i}(\vec{r},t)+\Nabla_{\vec{r}}\!\cdot\!\hat{\vec{J}}^{(i)}(\vec{r},t)=0
\label{eq:AcO}%
\end{equation}
with the local currents $\hat{\vec{J}}^{(i)}(\vec{r},t)$ and $i\in\{1,\dotsc,n_{\mathrm{c}}\}$.
Correspondingly, the dynamics of the averaged conserved relevant variables is given by  
\begin{equation}
\dot{a}^{\mathrm{c}}_{i}(\vec{r},t)+\Nabla_{\vec{r}}\!\cdot\!\vec{J}^{(i)}(\vec{r},t)=0 \;.
\label{eq:Ac}%
\end{equation}
On the other hand, it is assumed that the time-evolution of the $n_{\mathrm{n}}$ non-conserved relevant variables can be described by the 
balance equations
\begin{equation}
\dot{\hat{a}}^{\mathrm{n}}_{i}(\vec{r},t)+\hat{\Phi}^{(i)}(\vec{r},t)=0
\label{eq:AncO}%
\end{equation}
with the local quasi-currents $\hat{\Phi}^{(i)}(\vec{r},t)$ and $i\in\{1,\dotsc,n_{\mathrm{n}}\}$, while the 
averaged non-conserved relevant variables are described by 
\begin{equation}
\dot{a}^{\mathrm{n}}_{i}(\vec{r},t)+\Phi^{(i)}(\vec{r},t)=0 \;. 
\label{eq:Anc}%
\end{equation}
Both the currents $\vec{J}^{(i)}(\vec{r},t)$ and the quasi-currents $\Phi^{(i)}(\vec{r},t)$ can be 
decomposed into two different contributions \cite{PleinerB1996}:
{\allowdisplaybreaks
\begin{align}%
\begin{split}%
\vec{J}^{(i)}(\vec{r},t)=\vec{J}^{(i)}_{\mathrm{R}}(\vec{r},t)+\vec{J}^{(i)}_{\mathrm{D}}(\vec{r},t) \;,
\end{split}\label{eq:JDR}\\%
\begin{split}%
\Phi^{(i)}(\vec{r},t)=\Phi^{(i)}_{\mathrm{R}}(\vec{r},t)+\Phi^{(i)}_{\mathrm{D}}(\vec{r},t) \;.
\end{split}\label{eq:PhiDR}%
\end{align}}%
These are the reversible contributions denoted by the subscript \ZT{$\mathrm{R}$} and the dissipative contributions 
denoted by the subscript \ZT{$\mathrm{D}$}.
While the reversible contributions are isentropic, \ie, they are not associated with entropy production, the dissipative (reversible)
contributions increase the total entropy of the system.  

Examples for the general relevant variables $\hat{a}^{\mathrm{c}}_{i}(\vec{r},t)$ and $\hat{a}^{\mathrm{n}}_{i}(\vec{r},t)$ 
are a concentration field and a local polarization, respectively. There are only two possible relevant variables that have to be treated 
explicitly and cannot be taken into account by the general set $\hat{a}_{i}(\vec{r},t)$ of relevant variables. 
These special variables are the conserved generalized internal energy density 
$\hat{\varepsilon}(\vec{r},t)$ or its Legendre transforms and the non-conserved entropy density $\hat{\sigma}(\vec{r},t)$ 
(see Sec.\ \ref{subsec:NED}).\footnote{In fact, the internal energy density $\hat{\varepsilon}(\vec{r},t)$ 
has to be treated separately, since $\varepsilon(\vec{r},t)$ is a Legendre transform of the generalized Helmholtz 
free-energy density $f(\vec{r},t)$ that can be gradient expanded in terms of the other relevant variables. 
The entropy density $\hat{\sigma}(\vec{r},t)$, on the other hand, has to be treated separately, since it is not independent of the 
internal energy density $\hat{\varepsilon}(\vec{r},t)$. Both variables are related by the local formulation of the first law of thermodynamics.} 
Since the internal energy density is conserved, the transport equations for $\hat{\varepsilon}(\vec{r},t)$ and $\varepsilon(\vec{r},t)$ 
can be written as
{\allowdisplaybreaks
\begin{align}%
\begin{split}%
\dot{\hat{\varepsilon}}(\vec{r},t)+\Nabla_{\vec{r}}\!\cdot\!\hat{\vec{J}}^{\varepsilon}(\vec{r},t)=0 \;,
\end{split}\label{eq:eI}\\%
\begin{split}%
\dot{\varepsilon}(\vec{r},t)+\Nabla_{\vec{r}}\!\cdot\!\vec{J}^{\varepsilon}(\vec{r},t)=0
\end{split}\label{eq:eII}%
\end{align}}%
with the internal energy currents $\hat{\vec{J}}^{\varepsilon}(\vec{r},t)$ and 
$\vec{J}^{\varepsilon}(\vec{r},t)=\vec{J}^{\varepsilon}_{\mathrm{R}}(\vec{r},t)+\vec{J}^{\varepsilon}_{\mathrm{D}}(\vec{r},t)$.
The entropy density, in contrast, is non-conserved and the transport equations for 
$\hat{\sigma}(\vec{r},t)$ and $\sigma(\vec{r},t)$ are 
{\allowdisplaybreaks
\begin{align}%
\begin{split}%
\dot{\hat{\sigma}}(\vec{r},t)+\hat{\Phi}^{\sigma}(\vec{r},t)=0 \;,
\end{split}\label{eq:sI}\\%
\begin{split}%
\dot{\sigma}(\vec{r},t)+\Phi^{\sigma}(\vec{r},t)=0
\end{split}\label{eq:sII}%
\end{align}}%
with the entropy quasi-currents $\hat{\Phi}^{\sigma}(\vec{r},t)$ and 
$\Phi^{\sigma}(\vec{r},t)=\Phi^{\sigma}_{\mathrm{R}}(\vec{r},t)+\Phi^{\sigma}_{\mathrm{D}}(\vec{r},t)$.

\subsection{\label{subsec:DF}Dissipation functionals}
If the considered system is in local thermodynamic equilibrium so that the local formulation of the first and 
second law of thermodynamics hold, linear irreversible thermodynamics \cite{MartinPP1972,deGrootM1984,Reichl1998} 
can be applied \cite{PleinerB1996}.
This useful framework is especially applicable for all passive systems and states the existence of a dissipation functional $\mathfrak{R}$ 
from which the dissipative currents $\vec{J}^{(i)}_{\mathrm{D}}(\vec{r},t)$ and quasi-currents $\Phi^{(i)}_{\mathrm{D}}(\vec{r},t)$ 
of the averaged relevant variables $a_{i}(\vec{r},t)$ and the entropy density $\sigma(\vec{r},t)$ can be derived. 
The dissipation functional 
\begin{equation}
\mathfrak{R}=\!\int_{\R^{3}}\!\!\!\!\!\:\!\dif^{3}r\, \mathfrak{r}(\vec{r},t)
\label{eq:DF}%
\end{equation}
describes the total amount of energy that is dissipated per time in the considered system. 
Its integrand, the dissipation function $\mathfrak{r}(\vec{r},t)$, is therefore positive semi-definite as a 
consequence of the second law of thermodynamics \cite{Forster1990,PleinerB1996}: $\mathfrak{r}(\vec{r},t)\geqslant 0$.\footnote{The condition 
$\mathfrak{r}(\vec{r},t)\geqslant 0$ is a necessary prerequisite for dynamical stability of the considered system \cite{PleinerB1996}. 
For entirely reversible processes, $\mathfrak{r}(\vec{r},t)$ is zero, while it is positive for all other (dissipative) processes.}
It is nonlinear in the thermodynamic variables and quadratic in the \textit{thermodynamic forces} that are defined as
\begin{equation}
\vec{a}^{\mathrm{c}\sharp}_{i}=-\Nabla_{\vec{r}}\:\! a^{\mathrm{c}\natural}_{i}\;,\qquad
a^{\mathrm{n}\sharp}_{i}=a^{\mathrm{n}\natural}_{i}
\label{eq:TDF}%
\end{equation}
and correspond to the averaged relevant variables $a_{i}(\vec{r},t)$.
Again, the internal energy density $\varepsilon(\vec{r},t)$ and the entropy density $\sigma(\vec{r},t)$ have to be treated separately.
While a thermodynamic force corresponding to the internal energy density does not exist, the thermodynamic force $\sigma^{\sharp}(\vec{r},t)$ 
corresponding to the entropy density is defined further below by Eq.\ \eqref{eq:Jsigma}.
For a given dissipation functional, the dissipative currents and quasi-currents of the averaged relevant variables 
follow directly by functional differentiation with respect to the thermodynamic forces \cite{PleinerB1996}:
\begin{equation}
\vec{J}^{(i)}_{\mathrm{D}}=\Fdif{\mathfrak{R}}{\vec{a}^{\mathrm{c}\sharp}_{i}} \;,\qquad 
\Phi^{(i)}_{\mathrm{D}}=\Fdif{\mathfrak{R}}{a^{\mathrm{n}\sharp}_{i}} \;.
\label{eq:JPhi}%
\end{equation}
The dissipative energy current $\vec{J}^{\varepsilon}_{\mathrm{D}}(\vec{r},t)$ and the 
dissipative entropy quasi-current $\Phi^{\sigma}_{\mathrm{D}}(\vec{r},t)$ have to be derived differently 
(see Sec.\ \ref{subsec:NED} further below). 

For systems far from thermodynamic equilibrium, linear irreversible thermodynamics cannot be applied and 
a dissipation functional is in general not known \cite{Reichl1998}. 
This is the case for active systems like, for example, lasers, amplifiers, and biological systems. 
Indeed there is a generalization of the dissipation functional -- the generalized Ljapunov functional -- 
to systems that can also be far from thermodynamic equilibrium, but a Ljapunov functional does not exist in general. 
The criteria for the existence of a Ljapunov functional for a particular system 
are characterized by potential conditions \cite{GrahamH1971a,GrahamH1971b,Risken1972,Risken1996}.

\section{\label{sec:ED}The entropy density as a relevant variable}
The entropy density $\sigma(\vec{r},t)$ is a special (non-conserved) relevant variable that has not yet been considered in the context of 
EDDFT \cite{WittkowskiLB2012}. 
Although EDDFT could in principle be generalized to include also non-conserved variables, the entropy density cannot be incorporated into 
EDDFT directly. It has instead to be considered separately, since there is no corresponding operator $\hat{\sigma}(\vec{r},t)$ known for the 
entropy density that would be required to incorporate the entropy density into EDDFT using the MZFT \cite{WittkowskiLB2012}.    
To avoid this problem, we assume the applicability of the local formulation of the first law of thermodynamics for operators and express the 
unknown operator for the entropy density in terms of the known operators of the other relevant variables. We thus construct an expression 
for the entropy density operator that is applicable in the framework of linear irreversible thermodynamics. 
On this basis, we present the derivation of dynamical equations for the entropy density $\sigma(\vec{r},t)$ 
as well as for its time correlation function $C_{\sigma\sigma}(\vec{r},\rs\!,t,t')$.

\subsection{\label{subsec:NED}Non-equilibrium dynamics}
Since the entropy density $\sigma(\vec{r},t)$ is not conserved, its dynamics has to be described by the balance equation \eqref{eq:sII}, 
where the entropy quasi-current 
\begin{equation}
\Phi^{\sigma}(\vec{r},t)=\Nabla_{\vec{r}}\!\cdot\!\vec{J}^{\sigma}(\vec{r},t) - Q^{\sigma}(\vec{r},t) 
\label{eq:Phis}%
\end{equation}
can be decomposed into a divergence term with the entropy current 
$\vec{J}^{\sigma}(\vec{r},t)=\vec{J}^{\sigma}_{\mathrm{R}}(\vec{r},t)+\vec{J}^{\sigma}_{\mathrm{D}}(\vec{r},t)$ describing 
the transport of entropy and a source term $Q^{\sigma}(\vec{r},t)$ denoting the production of entropy.\footnote{The dissipative 
entropy quasi-current is therefore $\Phi^{\sigma}_{\mathrm{D}}(\vec{r},t)=\Nabla_{\vec{r}}\!\cdot\!\vec{J}^{\sigma}_{\mathrm{D}}(\vec{r},t) 
- Q^{\sigma}(\vec{r},t)$.}
The entropy production 
\begin{equation}
Q^{\sigma}(\vec{r},t)=\frac{2\:\!\mathfrak{r}(\vec{r},t)}{T(\vec{r},t)}
\label{eq:Q}%
\end{equation}
in turn can be expressed in terms of the dissipation function $\mathfrak{r}(\vec{r},t)$ and the absolute local temperature $T(\vec{r},t)$. 
In order to derive explicit microscopic expressions for $\vec{J}^{\sigma}(\vec{r},t)$ and $Q^{\sigma}(\vec{r},t)$, 
we make use of the local formulation of the first law of thermodynamics\footnote{Einstein's sum convention is assumed in the following. 
Equation \eqref{eq:HSI} makes the internal energy density $\varepsilon(\vec{r},t)$ and the entropy density $\sigma(\vec{r},t)$ 
dependent on each other and is the reason, why $\varepsilon(\vec{r},t)$ and $\sigma(\vec{r},t)$ have to be treated separately from 
the other relevant variables $a_{i}(\vec{r},t)$ (see Sec.\ \ref{subsec:TE}).} \cite{PleinerB1996}
\begin{equation}
\dif\varepsilon = T\dif\sigma + a^{\mathrm{c}\natural}_{i}\dif a^{\mathrm{c}}_{i} + a^{\mathrm{n}\natural}_{i}\dif a^{\mathrm{n}}_{i}
\label{eq:HSI}%
\end{equation}
with the generalized internal energy density $\varepsilon(\vec{r},t)$, 
which is related to the generalized Helmholtz free-energy density $f(\vec{r},t)$ by 
\begin{equation}
\dif f = \dif\varepsilon - T\dif\sigma - \sigma\dif T \;,
\end{equation}
and additional conserved and non-conserved classical thermodynamic variables 
$a^{\mathrm{c}}_{i}(\vec{r},t)$ and $a^{\mathrm{n}}_{i}(\vec{r},t)$, respectively.\footnote{Notice that the thermodynamic conjugate of the 
entropy density is the local absolute temperature: 
$\sigma^{\natural}(\vec{r},t)=\partial \varepsilon(\vec{r},t)/\partial\sigma(\vec{r},t)=T(\vec{r},t)$.} 
Equation \eqref{eq:HSI} can be rearranged into 
\begin{equation}
\dot{\sigma}=\frac{1}{T}\:\!\dot{\varepsilon}-\frac{a^{\mathrm{c}\natural}_{i}}{T}\:\!\dot{a}^{\mathrm{c}}_{i}
-\frac{a^{\mathrm{n}\natural}_{i}}{T}\:\!\dot{a}^{\mathrm{n}}_{i}
\label{eq:sigma0}%
\end{equation}
providing a dynamical equation for the entropy density in terms of the time derivatives\footnote{Throughout the whole paper, 
a dot $\dot{\,\,\,}$ denotes a derivation with respect to time $t$. This applies also to time derivatives of correlation functions with
two time variables $t$ and $t'$ further below, where the dot always means a derivation with respect to $t$ and not to $t'$.} 
of the internal energy density and the other relevant variables.
Using the transport equations \eqref{eq:Ac} and \eqref{eq:Anc} for the general relevant variables $a^{\mathrm{c}}_{i}(\vec{r},t)$ and 
$a^{\mathrm{n}}_{i}(\vec{r},t)$, respectively, \eqref{eq:eII} for the internal energy density $\varepsilon(\vec{r},t)$, 
and \eqref{eq:sII} for the entropy density $\sigma(\vec{r},t)$, Eq.\ \eqref{eq:sigma0} can be transformed into the 
\textit{balance equation for the entropy density} \eqref{eq:sII} with the decomposition \eqref{eq:Phis}, 
the dissipative entropy current
\begin{equation}
\vec{J}^{\sigma}_{\mathrm{D}}(\vec{r},t)=\frac{1}{T}\:\!\vec{J}^{\varepsilon}_{\mathrm{D}}
-\frac{a^{\mathrm{c}\natural}_{i}}{T}\:\!\vec{J}^{(i)}_{\mathrm{D}} \;,
\label{eq:sigmaII}%
\end{equation}
and the entropy production 
\begin{equation}
Q^{\sigma}(\vec{r},t)=-\frac{1}{T}\vec{J}^{\sigma}_{\mathrm{D}}\!\cdot\!\Nabla_{\vec{r}}T
-\frac{1}{T}\vec{J}^{(i)}_{\mathrm{D}}\!\cdot\!\Nabla_{\vec{r}}a^{\mathrm{c}\natural}_{i}
+\frac{1}{T}\Phi^{(i)}_{\mathrm{D}}a^{\mathrm{n}\natural}_{i} \;. 
\label{eq:sigmaIII}%
\end{equation}
Notice that only the dissipative currents and qua\-si-cur\-rents contribute to the entropy production 
so that the following equation must hold:
\begin{equation}
\Nabla_{\vec{r}}\!\cdot\!\vec{J}^{\varepsilon}_{\mathrm{R}} = 
T\:\!\Nabla_{\vec{r}}\!\cdot\!\vec{J}^{\sigma}_{\mathrm{R}}  
+a^{\mathrm{c}\natural}_{i}\Nabla_{\vec{r}}\!\cdot\!\vec{J}^{(i)}_{\mathrm{R}}
+a^{\mathrm{n}\natural}_{i}\Phi^{(i)}_{\mathrm{R}} \;.
\end{equation}
Combining Eqs.\ \eqref{eq:Q} and \eqref{eq:sigmaIII} now leads to the desired microscopic expression 
for the \textit{dissipation function}:  
\begin{equation}
2\:\!\mathfrak{r}(\vec{r},t)=-\vec{J}^{\sigma}_{\mathrm{D}}\!\cdot\!\Nabla_{\vec{r}}T
-\vec{J}^{(i)}_{\mathrm{D}}\!\cdot\!\Nabla_{\vec{r}}a^{\mathrm{c}\natural}_{i}
+\Phi^{(i)}_{\mathrm{D}}a^{\mathrm{n}\natural}_{i} \;. 
\label{eq:r}%
\end{equation}
Since the dissipation function $\mathfrak{r}(\vec{r},t)$ has to be positive semi-definite, Eq.\ \eqref{eq:r} implies that the dissipative 
currents and quasi-currents are functions of the thermodynamic forces \eqref{eq:TDF}.
In fact, these functions are assumed to be linear and homogeneous \cite{PleinerB1996}.
With this assumption, is is obvious that the dissipative currents $\vec{J}^{(i)}_{\mathrm{D}}(\vec{r},t)$ and 
quasi-currents $\Phi^{(i)}_{\mathrm{D}}(\vec{r},t)$ corresponding to the conserved relevant variables 
$a^{\mathrm{c}}_{i}(\vec{r},t)$ and the non-conserved relevant variables $a^{\mathrm{n}}_{i}(\vec{r},t)$, respectively, 
can be derived from a given dissipation functional $\mathfrak{R}$ using Eqs.\ \eqref{eq:JPhi}.
Analogously, the dissipative entropy current $\vec{J}^{\sigma}_{\mathrm{D}}(\vec{r},t)$ can be derived:
\begin{equation}%
\sigma^{\natural}=T \;,\qquad
\vec{\sigma}^{\sharp}= -\Nabla_{\vec{r}}\:\! T \;,\qquad
\vec{J}^{\sigma}_{\mathrm{D}}=\Fdif{\mathfrak{R}}{\vec{\sigma}^{\sharp}} \;.
\label{eq:Jsigma}%
\end{equation}%
The dissipative energy current $\vec{J}^{\varepsilon}_{\mathrm{D}}(\vec{r},t)$ follows then from Eq.\ \eqref{eq:sigmaII}.

Equation \eqref{eq:sII} together with Eqs.\ \eqref{eq:Phis} and \eqref{eq:sigmaII}-\eqref{eq:r} embody a microscopic expression for the 
dynamical equation of the entropy density. This microscopic equation expresses the entropy quasi-current and the dissipation functional 
in terms of the (microscopic) currents and quasi-currents of the other relevant variables and constitutes the first main result of this paper.
It is noteworthy that the equations in this subsection closely resemble the corresponding hydrodynamic equations \cite{PleinerB1996}.

\subsection{Equilibrium correlations}
We now consider the dynamics of the entropy time correlation function $C_{\sigma\sigma}(\vec{r},\rs\!,t,t')$.
In the linear regime near equilibrium, the equilibrium time correlation function (Kubo function \cite{Forster1990}) 
of two variables $\hat{X}(\vec{r},t)$ and $\hat{Y}(\vec{r},t)$ is defined as \cite{Forster1990,WittkowskiLB2012}
\begin{equation}
C_{XY}(\vec{r},\rs\!,t,t')=\langle\Delta\hat{X}^{\mathrm{eq}}(\vec{r},t) | 
\Delta\hat{Y}^{\mathrm{eq}}(\rs\!,t')\rangle_{\mathrm{eq}} \;.
\label{eq:Cij}%
\end{equation}
Here, the letters \ZT{eq} denote equilibrium quantities, $\langle\cdot|\cdot\rangle_{\mathrm{eq}}$ is Mori's scalar product \cite{WittkowskiLB2012}, 
and $\Delta\hat{X}^{\mathrm{eq}}(\vec{r},t)=\hat{X}(\vec{r},t)-X^{\mathrm{eq}}(\vec{r})$ with 
$X^{\mathrm{eq}}(\vec{r})=\langle\hat{X}(\vec{r},t)\rangle_{\mathrm{eq}}$ 
are the equilibrium fluctuations of the variable $\hat{X}(\vec{r},t)$. 
Since they are associated with equilibrium fluctuations, the time correlation functions \eqref{eq:Cij} are translationally invariant 
with respect to time: $C_{XY}(\vec{r},\rs\!,t,t')=C_{XY}(\vec{r},\rs\!,t-t')$.
For the derivation of the dynamics of $C_{\sigma\sigma}(\vec{r},\rs\!,t,t')$, we need an expression for the operator $\hat{\sigma}(\vec{r},t)$ 
of the entropy density, but such an operator is not known in general. 
Therefore, we assume that the local formulation of the first law of thermodynamics \eqref{eq:HSI} 
also holds for the corresponding operators \cite{KadanoffM1963,HohenbergM1965,Forster1974,BrandDG1979,Forster1990}.  
This leads to the equation\footnote{Although Eq.\ \eqref{eq:HSIO} provides an expression for the entropy operator $\hat{\sigma}(\vec{r},t)$, 
this variable cannot be incorporated into EDDFT in the usual way, since the MZFT requires a set of \textit{independent} relevant variables 
\cite{WittkowskiLB2012}.} 
\begin{equation}
\dif\hat{\varepsilon} = T\dif\hat{\sigma} + a^{\natural}_{i}\dif\hat{a}_{i} 
\label{eq:HSIO}%
\end{equation}
that expresses the entropy density operator $\hat{\sigma}(\vec{r},t)$ in terms of the internal energy density operator $\hat{\varepsilon}(\vec{r},t)$ 
and the operators $\hat{a}_{i}(\vec{r},t)$ for the other relevant variables that are known.  
Notice that the relevant variables $\hat{a}_{i}\in\{\hat{a}^{\mathrm{c}}_{i},\hat{a}^{\mathrm{n}}_{i}\}$ are not distinguished into 
conserved and non-conserved variables in this paragraph. 
Equation \eqref{eq:HSIO} can now be rearranged into 
\begin{equation}
\dot{\hat{\sigma}}=\frac{1}{T}\:\!\dot{\hat{\varepsilon}}-\frac{a^{\natural}_{i}}{T}\:\!\dot{\hat{a}}_{i}
\end{equation}
providing a dynamical equation for the entropy density operator.
Using the definition \eqref{eq:Cij} of the time correlation functions and their symmetry properties described in Ref.\ \cite{Forster1990}, 
relations between the time-derived correlation functions $\dot{C}_{XY}(\vec{r},\rs\!,t,t')$ with $X,Y\in\{\sigma,\varepsilon,a_{i}\}$ 
can be derived. 
A combination of these relations leads to the following 
\textit{transport equation for the entropy time correlation function}:    
\begin{equation}
\begin{split}%
\dot{C}_{\sigma\sigma}(\vec{r},\rs\!,t,t')=\,&\frac{1}{T(\vec{r},t)T(\rs\!,t')}\:\!\dot{C}_{\varepsilon\varepsilon}(\vec{r},\rs\!,t,t')\\
&-\frac{a^{\natural}_{i}(\vec{r},t)}{T(\vec{r},t)T(\rs\!,t')}\:\!\dot{C}_{a_{i}\varepsilon}(\vec{r},\rs\!,t,t')\\
&-\frac{a^{\natural}_{i}(\rs\!,t')}{T(\vec{r},t)T(\rs\!,t')}\:\!\dot{C}_{\varepsilon a_{i}}(\vec{r},\rs\!,t,t')\\
&+\frac{a^{\natural}_{i}(\vec{r},t)a^{\natural}_{j}(\rs\!,t')}{T(\vec{r},t)T(\rs\!,t')}\:\!\dot{C}_{a_{i}a_{j}}(\vec{r},\rs\!,t,t') \;.
\end{split}\raisetag{6.4em}%
\label{eq:Csigmasigma}%
\end{equation}
This equation constitutes the second main result of this paper, since it brings about a microscopic expression for the 
entropy time correlation function.

\subsection{Dissipation}
By the entropy balance equation \eqref{eq:sII} with the quasi-current \eqref{eq:Phis} and the entropy production \eqref{eq:Q}, 
the dissipation function \eqref{eq:r} is directly associated with dissipation. 
Also the time-derived correlation functions in Eq.\ \eqref{eq:Csigmasigma} can be related to dissipation, 
when the validity of the fluctuation-dissipation theorem \cite{CallenW1951,Forster1990} is assumed. 
For systems sufficiently close to thermodynamic equilibrium, the fluctuation-dissipation theorem implies the relation 
\begin{equation}
\dot{C}_{a_{i}a_{j}}(\vec{r},\rs\!,t,t')=\frac{2}{\ii\beta} \:\!\chi''_{ij}(\vec{r},\rs\!,t,t') 
\label{eq:FDTI}%
\end{equation}
or equivalently in Fourier space (see appendix \ref{app:FT})
\begin{equation}%
\widetilde{\chi}''_{ij}(\vec{r},\rs\!,\omega)=\frac{\beta}{2}\:\!\omega\:\! \widetilde{C}_{a_{i}a_{j}}(\vec{r},\rs\!,\omega) 
\label{eq:FDTII}%
\end{equation}%
between the time correlation function $C_{a_{i}a_{j}}(\vec{r},\rs\!,t,t')$ and the 
absorptive response function\footnote{See Refs.\ \cite{KadanoffM1963,Forster1990,ChaikinL1995} for the symmetry properties 
of the absorptive response function $\chi''_{ij}(\vec{r},\rs\!,t,t')$.} \cite{WittkowskiLB2012}
\begin{equation}
\chi''_{ij}(\vec{r},\rs\!,t,t')=\frac{1}{2\hbar}\langle[\hat{a}_{i}(\vec{r},t),\hat{a}_{j}(\rs\!,t')]\rangle_{\mathrm{eq}} 
\end{equation}
for any variables $\hat{a}_{i}(\vec{r},t)$ and $\hat{a}_{j}(\vec{r},t)$ including also the internal energy density $\hat{\varepsilon}(\vec{r},t)$ 
and the entropy density $\hat{\sigma}(\vec{r},t)$. 
In the following, the meaning of the dissipation functional $\mathfrak{R}$ and the 
absorptive response function $\chi''_{ij}(\vec{r},\rs\!,t,t')$ in the context of dissipation is considered in more detail.

\subsubsection{Non-equilibrium dynamics}
The total work $W(t)$ done on a system with the effective Hamiltonian $\hat{H}_{\mathrm{eff}}(t)$ is given by \cite{Forster1990,ChaikinL1995}
\begin{equation}
W(t)=\Tr(\rho(t)\hat{H}_{\mathrm{eff}}(t)) \;. 
\end{equation}
Its rate of change $\dot{W}(t)$ is the energy dissipated per time in this system and identical with the 
dissipation functional \cite{KadanoffM1963,Forster1990}
\begin{equation}
\begin{split}%
\mathfrak{R}&=\dot{W}(t)=\Tr(\rho(t)\dot{\hat{H}}_{\mathrm{eff}}(t)) \\
&=-\sum^{n}_{i=1}\int_{\R^{3}}\!\!\!\!\!\:\!\dif^{3}r\, a_{i}(\vec{r},t) \dot{a}^{\natural}_{i}(\vec{r},t) \;, 
\end{split}%
\label{eq:Rdiss}%
\end{equation}
where the averaged relevant variables $a_{i}(\vec{r},t)$ now include also the internal energy density $\varepsilon(\vec{r},t)$, 
but not the dependent entropy density $\sigma(\vec{r},t)$. 
Notice that Eq.\ \eqref{eq:Rdiss} can be understood as a generalization of Eq.\ (33) in Ref.\ \cite{EspanolL2009}. 
The time integral of Eq.\ \eqref{eq:Rdiss} is the total amount of dissipated energy   
\begin{equation}
W_{\mathrm{diss}}=\int_{\R}\!\!\hskip-0.5pt\dif t\,\:\!\mathfrak{R} 
=\sum^{n}_{i=1}\int_{\R}\!\!\hskip-0.5pt\dif t \!\int_{\R^{3}}\!\!\!\!\!\:\!\dif^{3}r\,\dot{a}_{i}(\vec{r},t) a^{\natural}_{i}(\vec{r},t) \;. 
\label{eq:WdissI}%
\end{equation}

\subsubsection{Equilibrium correlations}
Using the fluctuation-dissipation theorem \eqref{eq:FDTI}, the dissipated energy \eqref{eq:WdissI} can also be expressed 
in dependence of the time-derived correlation functions $\dot{C}_{a_{i}a_{j}}(\vec{r},\rs\!,t,t')$ or equivalently in dependence of the  
absorptive response functions $\chi''_{ij}(\vec{r},\rs\!,t,t')$. 
For this purpose, the symmetric equilibrium susceptibility matrix (complex response function) \cite{WittkowskiLB2012}
\begin{equation}
\begin{split}%
\chi_{ij}(\vec{r},\rs\!,t,t')=\Fdif{a_{i}(\vec{r},t)}{a^{\natural}_{j}(\rs\!,t')}\bigg\rvert_{\mathrm{eq}}
\end{split}%
\label{eq:chi_ij}%
\end{equation}
is used in order to express the averaged relevant variables $a_{i}(\vec{r},t)=a^{\mathrm{eq}}_{i}(\vec{r})+\Delta a_{i}(\vec{r},t)$ 
in terms of the corresponding thermodynamic conjugates $a^{\natural}_{i}(\vec{r},t)$\footnote{In the linear regime near equilibrium, 
Eq.\ \eqref{eq:a_chi} follows directly from a functional Taylor expansion of the variable $a_{i}(\vec{r},t)$ with respect to the 
thermodynamic conjugates $a^{\natural}_{i}(\vec{r},t)$ about the equilibrium state, in which the thermodynamic conjugates vanish.}: 
\begin{equation}
\Delta a_{i}(\vec{r},t)=\sum^{n}_{j=1}\int_{\R}\!\!\hskip-0.5pt\dif t' \!\int_{\R^{3}}\!\!\!\!\!\:\!\dif^{3}r'\,
\chi_{ij}(\vec{r},\rs\!,t,t')a^{\natural}_{j}(\rs\!,t') \;. 
\label{eq:a_chi}%
\end{equation}
Inserting this expression into Eq.\ \eqref{eq:WdissI} leads to 
\begin{equation}
\begin{split}%
W_{\mathrm{diss}}=\sum^{n}_{i,j=1}&\int_{\R}\!\!\hskip-0.5pt\dif t \!\int_{\R}\!\!\hskip-0.5pt\dif t'\! 
\!\int_{\R^{3}}\!\!\!\!\!\:\!\dif^{3}r\!\!\int_{\R^{3}}\!\!\!\!\!\:\!\dif^{3}r'\, \\
&\times a^{\natural}_{i}(\vec{r},t)\:\!\dot{\chi}_{ij}(\vec{r},\rs\!,t,t')\:\!a^{\natural}_{j}(\rs\!,t') \;. 
\end{split}%
\label{eq:WdissII}%
\end{equation}
In Fourier space, the complex response function \eqref{eq:chi_ij} can be decomposed as \cite{Forster1990,WittkowskiLB2012}
\begin{equation}
\widetilde{\chi}_{ij}(\vec{r},\rs\!,\omega) = \widetilde{\chi}'_{ij}(\vec{r},\rs\!,\omega) 
+ \ii\, \widetilde{\chi}''_{ij}(\vec{r},\rs\!,\omega)
\end{equation}
with the (Fourier transformed) reactive response function $\widetilde{\chi}'_{ij}(\vec{r},\rs\!,\omega)$ and 
absorptive response function $\widetilde{\chi}''_{ij}(\vec{r},\rs\!,\omega)$. 
Using this decomposition, the symmetry properties \cite{KadanoffM1963,Forster1990,ChaikinL1995} 
$\widetilde{\chi}'_{ji}(\rs\!,\vec{r},-\omega)=\widetilde{\chi}'_{ij}(\vec{r},\rs\!,\omega)$ and 
$-\widetilde{\chi}''_{ji}(\rs\!,\vec{r},-\omega)=\widetilde{\chi}''_{ij}(\vec{r},\rs\!,\omega)$
of the reactive and absorptive response functions, respectively, as well as Parseval's theorem (see appendix \ref{app:PT}),   
it can be shown that Eq.\ \eqref{eq:WdissII} is equivalent to 
\begin{equation}%
\begin{split}%
W_{\mathrm{diss}}=\frac{1}{2\pi}\sum^{n}_{i,j=1}&\int_{\R^{3}}\!\!\!\!\!\:\!\dif^{3}r\!\int_{\R^{3}}\!\!\!\!\!\:\!\dif^{3}r'\!\!
\int_{\R}\!\!\hskip-0.5pt\dif\omega\, \\
&\times \widetilde{a}^{\natural}_{i}(\vec{r},-\omega)\:\!\omega\:\!\widetilde{\chi}''_{ij}(\vec{r},\rs\!,\omega)
\widetilde{a}^{\natural}_{j}(\rs\!,\omega) \;. 
\end{split}\raisetag{3em}%
\label{eq:WdissIII}%
\end{equation}%
In accordance with the second law of thermodynamics, the dissipated energy $W_{\mathrm{diss}}$ must be non-negative 
in a stable system (see Sec.\ \ref{subsec:DF}).
Since this stability condition must also hold for any generalized Helmholtz free-energy functional $\mathcal{F}$ and hence for 
an arbitrary set of thermodynamic conjugates $\widetilde{a}^{\natural}_{i}(\vec{r},\omega)$,
the product\footnote{Notice that the sign of this expression depends on the definition of the Fourier transformation.} 
$\omega\:\!\widetilde{\chi}''_{ij}(\vec{r},\rs\!,\omega)$ has to be positive semi-definite for any $\omega$ \cite{Forster1990}. 
As a consequence of Eq.\ \eqref{eq:FDTII}, also $\widetilde{C}_{a_{i}a_{j}}(\vec{r},\rs\!,\omega)$ has to be positive semi-definite.

\section{\label{sec:EDDFT}Extended dynamical density functional theory}
So far, a general set of relevant conserved or non-conserved variables $\hat{a}_{i}(\vec{r},t)$ with general transport equations 
\eqref{eq:AcO} and \eqref{eq:AncO}, respectively, has been considered. 
In this section, we now focus on current EDDFT \cite{WittkowskiLB2012} with its conserved, real, and independent variables 
$\hat{a}_{i}(\vec{r},t)\equiv \hat{a}^{\mathrm{c}}_{i}(\vec{r},t)$ with $i\in\{1,\dotsc,n\}$ including the internal energy density 
and with the corresponding specified transport equations. 
We at first summarize the EDDFT equations and the associated transport equations for time correlation functions. 
Afterwards, we discuss the derivation of EDDFT from a dissipation functional and consider the hydrodynamic limit of EDDFT. 
These considerations are valid for general variables $\hat{a}_{i}(\vec{r},t)$ that have to be conserved, real, and independent, as usual in the 
context of EDDFT \cite{WittkowskiLB2012}. 
For more particular EDDFT equations with the concentrations $\hat{c}_{i}(\vec{r},t)$ and the energy density $\hat{\varepsilon}(\vec{r},t)$
as relevant variables, see Ref.\ \cite{WittkowskiLB2012}.

\subsection{\label{sec:EDDFTa}Non-equilibrium dynamics}
In its currently most general form, the \textit{EDDFT equations} are given by \cite{WittkowskiLB2012}
\begin{equation}
\dot{a}_{i}(\vec{r},t) + \Nabla_{\vec{r}}\!\cdot\!\vec{J}^{a_{i}}(\vec{r},t) = 0
\label{eq:EDDFTa}%
\end{equation}
with the currents $\vec{J}^{a_{i}}(\vec{r},t)=\vec{J}^{a_{i}}_{\mathrm{R}}(\vec{r},t)+\vec{J}^{a_{i}}_{\mathrm{D}}(\vec{r},t)$ and
{\allowdisplaybreaks
\begin{align}%
\begin{split}%
\vec{J}^{a_{i}}_{\mathrm{R}}(\vec{r},t)&=\Tr\!\big(\rho(t)\hat{\vec{J}}^{(i)}(\vec{r},0)\big) \;,
\end{split}\label{eq:EDDFTb}\\%
\begin{split}%
\vec{J}^{a_{i}}_{\mathrm{D}}(\vec{r},t)&=-\sum^{n}_{j=1}\int_{\R^{3}}\!\!\!\!\!\:\!\dif^{3}r'\,\beta D^{(ij)}(\vec{r},\rs\!,t)
\Nabla_{\rs}a^{\natural}_{j}(\rs\!,t) \;.
\end{split}\label{eq:EDDFTc}%
\end{align}}%
Here, $D^{(ij)}(\vec{r},\rs\!,t)$ denotes a diffusion tensor, for which explicit expressions can be found in Ref.\ \cite{WittkowskiLB2012}.
We note that $D^{(ij)}(\vec{r},\rs\!,t)$ in Eq.\ \eqref{eq:EDDFTc} contains 
all non-instantaneous contributions. These are mainly dissipative in nature. 
For certain systems such as, for example, nematic liquid crystals, there are also reversible 
contributions even in the hydrodynamic regime \cite{Forster1974,Forster1990}.

\subsection{\label{sec:EDDFTb}Equilibrium correlations}
In the linear regime near equilibrium, the corresponding transport equations for the equilibrium time correlation functions 
$C_{a_{i}a_{j}}(\vec{r},\rs\!,t)$\footnote{Here, we made use of the fact that the equilibrium time correlation functions 
$C_{a_{i}a_{j}}(\vec{r},\rs\!,t,t')$ are translationally invariant with respect to time so that $t'$ can be omitted.} 
are given by \cite{WittkowskiLB2012}
\begin{equation}
\dot{C}_{a_{i}a_{j}}(\vec{r},\rs\!,t) + \Nabla_{\vec{r}}\!\cdot\!\mathrm{J}^{(ij)}(\vec{r},\rs\!,t) = 0
\label{eq:EDDFTCFa}%
\end{equation}
with the total currents 
$\mathrm{J}^{(ij)}(\vec{r},\rs\!,t)=\mathrm{J}^{(ij)}_{\mathrm{R}}(\vec{r},\rs\!,t)+\mathrm{J}^{(ij)}_{\mathrm{D}}(\vec{r},\rs\!,t)$ 
and their contributions 
{\allowdisplaybreaks
\begin{align}%
\begin{split}%
\!\!\! \mathrm{J}^{(ij)}_{\mathrm{R}}(\vec{r},\rs\!,t) = -\sum^{n}_{k=1}\int_{\R^{3}}\!\!\!\!\!\:\!\dif^{3}r''\, 
\Omega^{(ik)}_{\mathrm{eq}}(\vec{r},\rss)\:\! C_{kj}(\rss\!,\rs\!,t) \,, \!\!\!
\end{split}\label{eq:EDDFTCFb}\\%
\begin{split}%
\!\!\! \mathrm{J}^{(ij)}_{\mathrm{D}}(\vec{r},\rs\!,t) = -\sum^{n}_{k=1}\int_{\R^{3}}\!\!\!\!\!\:\!\dif^{3}r''\, 
\Gamma^{(ik)}_{\mathrm{eq}}(\vec{r},\rss)\:\! C_{kj}(\rss\!,\rs\!,t) \,. \!\!\!
\end{split}\label{eq:EDDFTCFc}%
\end{align}}%
For explicit expressions for $\Omega^{(ij)}_{\mathrm{eq}}(\vec{r},\rs)$ and $\Gamma^{(ij)}_{\mathrm{eq}}(\vec{r},\rs)$, 
see Ref.\ \cite{WittkowskiLB2012}.

\subsection{\label{sec:EDDFTc}Dissipation}
The reversible currents \eqref{eq:EDDFTb} of EDDFT often vanish. 
This is in general the case, when all relevant variables $a_{i}(\vec{r},t)$ have the same time-reversal behavior \cite{Forster1990}.
Then only the dissipative currents \eqref{eq:EDDFTc} remain and the complete EDDFT equations \eqref{eq:EDDFTa} can be derived 
from a dissipation functional that is defined by Eqs.\ \eqref{eq:DF} and \eqref{eq:r}.
Notice that the internal energy density is treated separately from the other relevant variables in Eq.\ \eqref{eq:r}, 
while it is included in the relevant variables $a_{i}(\vec{r},t)$ in Eqs.\ \eqref{eq:EDDFTa}-\eqref{eq:EDDFTc}. 
For instance, the traditional DDFT of Marconi and Tarazona \cite{MarconiT1999,MarconiT2000,ArcherE2004} 
{\allowdisplaybreaks
\begin{gather}%
\begin{split}%
\dot{\rho}(\vec{r},t) + \Nabla_{\vec{r}}\!\cdot\!\vec{J}^{\rho}(\vec{r},t) = 0 \;,
\end{split}\\%
\begin{split}%
\vec{J}^{\rho}(\vec{r},t) = -\beta D_{0}\rho(\vec{r},t) \Nabla_{\vec{r}}\rho^{\natural}(\vec{r},t)
\end{split}%
\end{gather}}%
with the one-particle density $\rho(\vec{r},t)$ and the translational short-time diffusion coefficient $D_{0}$ 
can be derived formally from the dissipation functional $\mathfrak{R}$ with the dissipation function 
\begin{equation}
\mathfrak{r}_{\mathrm{MT}}(\vec{r},t) = \frac{1}{2}\beta D_{0}\rho(\vec{r},t) \big(\Nabla_{\vec{r}}\rho^{\natural}(\vec{r},t)\big)^{2} \;. 
\end{equation}
This reformulation of DDFT in terms of a dissipation functional constitutes an alternative representation of DDFT besides the usual 
DDFT equations. Such a representation by a dissipation functional is very advantageous, when some of the relevant variables 
(in this example the one-particle density $\rho(\vec{r},t)$) have to be parametrized by other order-parameter fields and 
dynamical equations for these parameterizing order-parameter fields are needed. 
The derivation of PFC models from DDFT, for example, involves such a parametrization of the one-particle density.
While this is pretty straightforward for isotropic systems with a purely translational particle density \cite{ElderPBSG2007,vanTeeffelenBVL2009},
the derivation of the dynamical equations is getting considerably more complicated for additional orientational degrees of freedom both in 
two \cite{Loewen2010,WittkowskiLB2011,WittkowskiLB2011b} and three \cite{WittkowskiLB2010} spatial dimensions
(for a recent review see \cite{EmmerichEtAl2012}). The usage of a dissipation functional instead of a DDFT equation considerably simplifies 
the situation.

\subsection{Special cases of the EDDFT equations}
In this subsection, we consider two special cases of the EDDFT equations \eqref{eq:EDDFTa}-\eqref{eq:EDDFTc} in order to demonstrate
their applicability and to compare them to equations that are known from the literature. 
These special cases are the dynamical equations for non-isothermal colloidal suspensions and the hydrodynamic limit of the EDDFT equations.

\subsubsection{Non-isothermal colloidal suspensions}
We consider a non-isothermal colloidal suspension, whose state is described by a given generalized Helmholtz free-energy functional 
$\mathcal{F}[c,T]$ in terms of two relevant variables, the (conserved) concentration field $c(\vec{r},t)$ 
and the (non-conserved) temperature field $T(\vec{r},t)$.
Such a functional can, for example, be obtained from static classical density functional theory (DFT) \cite{Oxtoby1991,Singh1991,Loewen1994a}, 
when a local approximation is applied to the DFT functional $\mathcal{F}(T,[c])$ with the spatially homogeneous temperature parameter $T$.
Instead of the temperature field $T(\vec{r},t)$, also the internal energy density $\varepsilon(\vec{r},t)$ or the entropy density $\sigma(\vec{r},t)$ 
lead -- in combination with the concentration field $c(\vec{r},t)$ -- to a set of two relevant variables that are appropriate to describe the 
considered system.  
Here, we choose $c(\vec{r},t)$ and $\sigma(\vec{r},t)$ as the relevant variables and the generalized internal energy functional 
\begin{equation}
\mathcal{E}[c,\sigma]=\mathcal{F}[c,T]+\int_{\R^{3}}\!\!\!\!\!\:\!\dif^{3}r\,T(\vec{r},t)\sigma(\vec{r},t) 
\end{equation}
as the corresponding thermodynamic functional. The thermodynamic conjugates of the relevant variables are thus defined as  
$c^{\natural}(\vec{r},t)=\delta\mathcal{E}/\delta c(\vec{r},t)$ and $\sigma^{\natural}(\vec{r},t)=\delta\mathcal{E}/\delta \sigma(\vec{r},t)$ 
in this section.
Since $c(\vec{r},t)$, $\varepsilon(\vec{r},t)$, and $\sigma(\vec{r},t)$ are even under parity inversion and time reversal, 
there are no reversible currents \cite{WittkowskiLB2012}:  
$\vec{J}^{c}_{\mathrm{R}}(\vec{r},t)=\vec{J}^{\varepsilon}_{\mathrm{R}}(\vec{r},t)=\vec{J}^{\sigma}_{\mathrm{R}}(\vec{r},t)=\vec{0}$.
Hence, the dynamics of such a non-isothermal colloidal suspension is described by 
{\allowdisplaybreaks
\begin{align}%
\begin{split}%
\dot{c}(\vec{r},t) + \Nabla_{\vec{r}}\!\cdot\!\vec{J}^{c}_{\mathrm{D}}(\vec{r},t) = 0 \;,
\end{split}\label{eq:NICSa}\\%
\begin{split}%
\dot{\sigma}(\vec{r},t) + 
\Nabla_{\vec{r}}\!\cdot\!\vec{J}^{\sigma}_{\mathrm{D}}(\vec{r},t) = Q^{\sigma}(\vec{r},t) \;.
\end{split}\label{eq:NICSb}%
\end{align}}%
These dynamical equations follow directly from Eqs.\ \eqref{eq:EDDFTa}, \eqref{eq:sII}, and \eqref{eq:Phis}.
The dissipative currents in Eqs.\ \eqref{eq:NICSa} and \eqref{eq:NICSb} result from Eq.\ \eqref{eq:EDDFTc} and are given by 
{\allowdisplaybreaks
\begin{align}%
\begin{split}%
\vec{J}^{c}_{\mathrm{D}}(\vec{r},t)=&-\int_{\R^{3}}\!\!\!\!\!\:\!\dif^{3}r'\,\beta D^{(cc)}(\vec{r},\rs\!,t)
\Nabla_{\rs}c^{\natural}(\rs\!,t) \\
&-\int_{\R^{3}}\!\!\!\!\!\:\!\dif^{3}r'\,\beta D^{(c\sigma)}(\vec{r},\rs\!,t)
\Nabla_{\rs}\sigma^{\natural}(\rs\!,t) \;,
\end{split}\label{eq:NICSc}\\%
\begin{split}%
\vec{J}^{\sigma}_{\mathrm{D}}(\vec{r},t)=&-\int_{\R^{3}}\!\!\!\!\!\:\!\dif^{3}r'\,\beta D^{(\sigma c)}(\vec{r},\rs\!,t)
\Nabla_{\rs}c^{\natural}(\rs\!,t) \\
&-\int_{\R^{3}}\!\!\!\!\!\:\!\dif^{3}r'\,\beta D^{(\sigma\sigma)}(\vec{r},\rs\!,t)
\Nabla_{\rs}\sigma^{\natural}(\rs\!,t) \;.
\end{split}\label{eq:NICSd}%
\end{align}}%
Using these expressions for the dissipative currents $\vec{J}^{c}_{\mathrm{D}}(\vec{r},t)$ and $\vec{J}^{\sigma}_{\mathrm{D}}(\vec{r},t)$, 
an explicit equation for the dissipative internal energy current 
$\vec{J}^{\varepsilon}_{\mathrm{D}}(\vec{r},t)$ can be derived from Eq.\ \eqref{eq:sigmaII}. 
This explicit equation is given by 
\begin{equation}
\begin{split}%
\vec{J}^{\varepsilon}_{\mathrm{D}}(\vec{r},t)=&-\int_{\R^{3}}\!\!\!\!\!\:\!\dif^{3}r'\,\beta D^{(\varepsilon c)}(\vec{r},\rs\!,t)
\Nabla_{\rs}c^{\natural}(\rs\!,t) \\
&-\int_{\R^{3}}\!\!\!\!\!\:\!\dif^{3}r'\,\beta D^{(\varepsilon\sigma)}(\vec{r},\rs\!,t)
\Nabla_{\rs}\sigma^{\natural}(\rs\!,t) 
\end{split}%
\label{eq:NICSe}%
\end{equation}
with the diffusion tensors 
{\allowdisplaybreaks
\begin{align}%
\begin{split}%
D^{(\varepsilon c)}(\vec{r},\rs\!,t) &= c^{\natural}(\vec{r},t) D^{(cc)}(\vec{r},\rs\!,t) \\
&\quad\:\!+ T(\vec{r},t) D^{(\sigma c)}(\vec{r},\rs\!,t) \;,
\end{split}\label{eq:Depsilonc}\\%
\begin{split}%
D^{(\varepsilon\sigma)}(\vec{r},\rs\!,t) &= c^{\natural}(\vec{r},t) D^{(c\sigma)}(\vec{r},\rs\!,t) \\
&\quad\:\!+ T(\vec{r},t) D^{(\sigma\sigma)}(\vec{r},\rs\!,t) \;.
\end{split}\label{eq:Depsilonsigma}%
\end{align}}%

\subsubsection{Hydrodynamic limit}
All expressions presented in this paper so far are beyond the scope of hydrodynamics and applicable 
also for non-hydrodynamic wave vectors $\vec{k}$ and frequencies $\omega$. 
In the hydrodynamic limit ($\vec{k}\to\vec{0}$, $\omega\to 0$), the transport matrices become constant and the transport equations 
simplify considerably. In particular, the diffusion tensor $D^{(ij)}(\vec{r},\rs\!,t)$ in Eq.\ \eqref{eq:EDDFTc} and the 
transport matrices $\Omega^{(ij)}_{\mathrm{eq}}(\vec{r},\rs)$ and $\Gamma^{(ij)}_{\mathrm{eq}}(\vec{r},\rs)$ in 
Eqs.\ \eqref{eq:EDDFTCFb} and \eqref{eq:EDDFTCFc}, respectively, are constant in the hydrodynamic limit 
(see Ref.\ \cite{WittkowskiLB2012} for details).

\section{\label{sec:conclusions}Conclusions and perspectives}
For a general set of independent relevant variables, we derived microscopic transport equations for the entropy density 
[see Eqs.\ \eqref{eq:sII}, \eqref{eq:Phis}, \eqref{eq:sigmaII}, and \eqref{eq:sigmaIII}] as well as for the entropy time correlation function 
[see Eq.\ \eqref{eq:Csigmasigma}] in the framework of linear irreversible thermodynamics. 
We thus complemented current EDDFT \cite{WittkowskiLB2012} with a balance equation for the entropy density 
that has previously not yet been considered in the context of EDDFT as a relevant variable.
We furthermore derived the microscopic expression \eqref{eq:r} defining a dissipation functional, from which the dissipative dynamics of EDDFT 
can be derived. This reformulation of EDDFT in terms of a dissipation functional complements the EDDFT equations and states that 
also DDFT \cite{MarconiT1999,MarconiT2000,ArcherE2004,EspanolL2009,WittkowskiL2011} can be derived from a dissipation functional. 
Our dissipation functional is especially useful, if some of the relevant variables have to be parametrized and dynamical equations 
for the parameterizing quantities are needed. This is in particular the case in the derivation of PFC models  
\cite{WittkowskiLB2010,WittkowskiLB2011,WittkowskiLB2011b,EmmerichEtAl2012} from EDDFT.
  
Our approach can be generalized to include macroscopic variables as well.
These are variables, which relax on a sufficiently long, but finite time scale
in the long wavelength limit.
Their combination with the hydrodynamic approach has been pioneered by 
Khalatnikov near the $\lambda$ transition in superfluid $^4$He \cite{Khalatnikov1989}.
In this case the degree of order, the modulus of the macroscopic wave function, 
becomes a slow macroscopic variable as this second-order phase transition is approached. 
Later the dynamics of the degree of order has also been introduced as a macroscopic variable 
near other phase transitions of second or weakly first order.

Another class of slow variables arises when there is an energy scale in the system, which
is much smaller than all other energy scales. A classical example of this case is superfluid 
$^3$He-A, where the magnetic dipole-dipole interaction is of order 
$2\times 10^{-7}\,\mathrm{K}$ (temperature associated with the appropriate energy) 
and thus very small compared to all other energy scales including the energy gap. 
As a consequence of the presence of the magnetic dipole-dipole interaction the magnetization density is no longer a truly conserved
hydrodynamic variable, but a rather long-lived macroscopic variable, 
whose dynamics can be incorporated into the hydrodynamics of superfluid $^3$He-A \cite{GrahamP1975,BrandDG1979}, 
where it turns out to be crucial for the understanding of the NMR spectra in the superfluid phase.

A class of macroscopic variables, which is particularly important
for complex fluids including polymers, colloids, liquid crystalline elastomers etc.\  
are strain fields associated with a transient network. In these cases the system 
reacts like a fluid below a certain frequency, the so-called Maxwell frequency,
and like a solid or glass above this frequency. Therefore, one has also explored 
the consequences of a transient strain field on the
macroscopic behavior by combining it with the hydrodynamic approach
\cite{BrandPR1990,TemmenPLB2000,TemmenPLB2001,PleinerLB2000,PleinerLB2002,PleinerLB2004}. 
It therefore appears like a natural next step to incorporate 
macroscopic variables such as a transient strain field into the EDDFT approach.

In contrast to hydrodynamic equations, our results are not restricted to the hydrodynamic regime ($\vec{k}\to\vec{0}$, $\omega\to 0$),   
but they are also applicable for larger wave vectors $\vec{k}$ and frequencies $\omega$.
However, our results that base on linear irreversible thermodynamics are restricted to systems that are sufficiently close to 
thermodynamic equilibrium, where the local formulation of the first and second law of thermodynamics are valid.
This is in general the case for all passive systems, but not for active systems.  
The generalization of our results to active systems that are far from thermodynamic equilibrium, where the entropy production can also be negative 
locally or in a subsystem and linear irreversible thermodynamics is not applicable, will therefore be an important task for the future.
In this context, especially the crucial question for the existence of an entropy density operator will have to be answered. 

Entropy has been considered in the context of DDFT in recent work of Espa\~{n}ol \cite{Espanol} and Schmidt \cite{Schmidt2011}. 
While in our work described in the present paper a generalized Helmholtz free-energy functional is chosen as an appropriate 
thermodynamic functional, the DDFT equations in the work of Espa\~{n}ol base on an entropy functional and do not include the entropy density 
as a thermodynamic variable \cite{Espanol}. 
As a further difference, the balance equation for the entropy density and the dissipation functional are not used in the approach of Espa\~{n}ol.
Schmidt, on the other hand, proposed DDFT equations for a one-particle density and an internal energy density on the basis of a 
generalized grand-canonical potential functional that depends functionally on the one-particle density and the entropy density and is minimized 
by both densities in thermodynamic equilibrium \cite{Schmidt2011}. 
The approach of Schmidt also includes a dissipation functional from which the dynamical equations for the one-particle density and the 
internal energy density can be derived, but -- opposed to our approach -- its microscopic justification is not emphasized and addressed.

\acknowledgments{We thank Pep Espa\~{n}ol for helpful discussions. This work was supported by the DFG within SPP 1296.}

\appendix
\section{Fourier transformation}
Since there are different definitions of the Fourier transformation in the literature, 
here we give the definition we used in the context of the work presented.
We further state a useful theorem related to the Fourier transformation.

\subsection{\label{app:FT}Definition} 
Within the definition we used, the \textit{Fourier transformation} of a time-dependent function $X(t)$ is given by 
\begin{equation}%
\begin{split}%
\widetilde{X}(\omega) &= \int_{\R}\!\!\hskip-0.5pt\dif t\, X(t) e^{\ii\omega t} \;, \\
X(t) &= \frac{1}{2\pi} \!\int_{\R}\!\!\hskip-0.5pt\dif \omega\,
\widetilde{X}(\omega) e^{-\ii\omega t} \\
\end{split}%
\end{equation}%
with $\omega\in\R$.

\subsection{\label{app:PT}Parseval's theorem}
If $X(t)$ and $Y(t)$ are two time-dependent square-integrable functions, \textit{Parseval's theorem} states
\begin{equation}
\int_{\R}\!\!\hskip-0.5pt\dif t\, X(t)\overline{Y}(t) 
=\frac{1}{2\pi}\int_{\R}\!\!\hskip-0.5pt\dif \omega\, \widetilde{X}(\omega)\overline{\widetilde{Y}}(\omega) \;, 
\end{equation}
where the bar $\overline{\,\cdot\,}$ denotes complex conjugation.

\bibliography{References}

\begin{thebibliography}{46}%
\makeatletter
\providecommand \@ifxundefined [1]{%
 \@ifx{#1\undefined}
}%
\providecommand \@ifnum [1]{%
 \ifnum #1\expandafter \@firstoftwo
 \else \expandafter \@secondoftwo
 \fi
}%
\providecommand \@ifx [1]{%
 \ifx #1\expandafter \@firstoftwo
 \else \expandafter \@secondoftwo
 \fi
}%
\providecommand \natexlab [1]{#1}%
\providecommand \enquote  [1]{``#1''}%
\providecommand \bibnamefont  [1]{#1}%
\providecommand \bibfnamefont [1]{#1}%
\providecommand \citenamefont [1]{#1}%
\providecommand \href@noop [0]{\@secondoftwo}%
\providecommand \href [0]{\begingroup \@sanitize@url \@href}%
\providecommand \@href[1]{\@@startlink{#1}\@@href}%
\providecommand \@@href[1]{\endgroup#1\@@endlink}%
\providecommand \@sanitize@url [0]{\catcode `\\12\catcode `\$12\catcode
  `\&12\catcode `\#12\catcode `\^12\catcode `\_12\catcode `\%12\relax}%
\providecommand \@@startlink[1]{}%
\providecommand \@@endlink[0]{}%
\providecommand \url  [0]{\begingroup\@sanitize@url \@url }%
\providecommand \@url [1]{\endgroup\@href {#1}{\urlprefix }}%
\providecommand \urlprefix  [0]{URL }%
\providecommand \Eprint [0]{\href }%
\providecommand \doibase [0]{http://dx.doi.org/}%
\providecommand \selectlanguage [0]{\@gobble}%
\providecommand \bibinfo  [0]{\@secondoftwo}%
\providecommand \bibfield  [0]{\@secondoftwo}%
\providecommand \translation [1]{[#1]}%
\providecommand \BibitemOpen [0]{}%
\providecommand \bibitemStop [0]{}%
\providecommand \bibitemNoStop [0]{.\EOS\space}%
\providecommand \EOS [0]{\spacefactor3000\relax}%
\providecommand \BibitemShut  [1]{\csname bibitem#1\endcsname}%
\let\auto@bib@innerbib\@empty
\bibitem [{\citenamefont {Reichl}(1998)}]{Reichl1998}%
  \BibitemOpen
  \bibfield  {author} {\bibinfo {author} {\bibfnamefont {L.~E.}\ \bibnamefont
  {Reichl}},\ }\href@noop {} {\emph {\bibinfo {title} {A Modern Course in
  Statistical Physics}}},\ \bibinfo {edition} {2nd}\ ed.\ (\bibinfo
  {publisher} {John Wiley \& Sons},\ \bibinfo {address} {New York},\ \bibinfo
  {year} {1998})\ p.\ \bibinfo {pages} {842}\BibitemShut {NoStop}%
\bibitem [{\citenamefont {Martin}\ \emph {et~al.}(1972)\citenamefont {Martin},
  \citenamefont {Parodi},\ and\ \citenamefont {Pershan}}]{MartinPP1972}%
  \BibitemOpen
  \bibfield  {author} {\bibinfo {author} {\bibfnamefont {P.~C.}\ \bibnamefont
  {Martin}}, \bibinfo {author} {\bibfnamefont {O.}~\bibnamefont {Parodi}}, \
  and\ \bibinfo {author} {\bibfnamefont {P.~S.}\ \bibnamefont {Pershan}},\
  }\href@noop {} {\bibfield  {journal} {\bibinfo  {journal} {Physical Review
  A}\ }\textbf {\bibinfo {volume} {6}},\ \bibinfo {pages} {2401} (\bibinfo
  {year} {1972})}\BibitemShut {NoStop}%
\bibitem [{\citenamefont {Groot}\ and\ \citenamefont
  {Mazur}(1984)}]{deGrootM1984}%
  \BibitemOpen
  \bibfield  {author} {\bibinfo {author} {\bibfnamefont {S.~R.~d.}\
  \bibnamefont {Groot}}\ and\ \bibinfo {author} {\bibfnamefont
  {P.}~\bibnamefont {Mazur}},\ }\href@noop {} {\emph {\bibinfo {title}
  {Non-Equilibrium Thermodynamics}}},\ \bibinfo {edition} {1st}\ ed.,\ Dover
  Books on Physics and Chemistry\ (\bibinfo  {publisher} {Dover Publications},\
  \bibinfo {address} {New York},\ \bibinfo {year} {1984})\ p.\ \bibinfo {pages}
  {544}\BibitemShut {NoStop}%
\bibitem [{\citenamefont {Mori}(1965)}]{Mori1965}%
  \BibitemOpen
  \bibfield  {author} {\bibinfo {author} {\bibfnamefont {H.}~\bibnamefont
  {Mori}},\ }\href@noop {} {\bibfield  {journal} {\bibinfo  {journal} {Progress
  of Theoretical Physics}\ }\textbf {\bibinfo {volume} {33}},\ \bibinfo {pages}
  {423} (\bibinfo {year} {1965})}\BibitemShut {NoStop}%
\bibitem [{\citenamefont {Zwanzig}\ and\ \citenamefont
  {Mountain}(1965)}]{ZwanzigM1965}%
  \BibitemOpen
  \bibfield  {author} {\bibinfo {author} {\bibfnamefont {R.}~\bibnamefont
  {Zwanzig}}\ and\ \bibinfo {author} {\bibfnamefont {R.~D.}\ \bibnamefont
  {Mountain}},\ }\href@noop {} {\bibfield  {journal} {\bibinfo  {journal}
  {Journal of Chemical Physics}\ }\textbf {\bibinfo {volume} {43}},\ \bibinfo
  {pages} {4464} (\bibinfo {year} {1965})}\BibitemShut {NoStop}%
\bibitem [{\citenamefont {Forster}(1974)}]{Forster1974}%
  \BibitemOpen
  \bibfield  {author} {\bibinfo {author} {\bibfnamefont {D.}~\bibnamefont
  {Forster}},\ }\href@noop {} {\bibfield  {journal} {\bibinfo  {journal}
  {Annals of Physics}\ }\textbf {\bibinfo {volume} {84}},\ \bibinfo {pages}
  {505} (\bibinfo {year} {1974})}\BibitemShut {NoStop}%
\bibitem [{\citenamefont {Grabert}(1982)}]{Grabert1982}%
  \BibitemOpen
  \bibfield  {author} {\bibinfo {author} {\bibfnamefont {H.}~\bibnamefont
  {Grabert}},\ }\href@noop {} {\emph {\bibinfo {title} {Projection Operator
  Techniques in Nonequilibrium Statistical Mechanics}}},\ \bibinfo {edition}
  {1st}\ ed.,\ \bibinfo {series} {Springer Tracts in Modern Physics},
  Vol.~\bibinfo {volume} {95}\ (\bibinfo  {publisher} {Springer},\ \bibinfo
  {address} {Berlin},\ \bibinfo {year} {1982})\ p.\ \bibinfo {pages}
  {166}\BibitemShut {NoStop}%
\bibitem [{\citenamefont {Forster}(1990)}]{Forster1990}%
  \BibitemOpen
  \bibfield  {author} {\bibinfo {author} {\bibfnamefont {D.}~\bibnamefont
  {Forster}},\ }\href@noop {} {\emph {\bibinfo {title} {Hydrodynamic
  Fluctuations, Broken Symmetry, and Correlation Functions}}},\ \bibinfo
  {edition} {1st}\ ed.,\ \bibinfo {series} {Advanced Book Classics},
  Vol.~\bibinfo {volume} {10}\ (\bibinfo  {publisher} {Perseus Books
  Publishing},\ \bibinfo {address} {New York},\ \bibinfo {year} {1990})\ p.\
  \bibinfo {pages} {352}\BibitemShut {NoStop}%
\bibitem [{\citenamefont {Dhont}(1996)}]{Dhont1996}%
  \BibitemOpen
  \bibfield  {author} {\bibinfo {author} {\bibfnamefont {J.~K.~G.}\
  \bibnamefont {Dhont}},\ }\href@noop {} {\emph {\bibinfo {title} {An
  Introduction to Dynamics of Colloids}}},\ \bibinfo {edition} {1st}\ ed.,\
  \bibinfo {series} {Studies in Interface Science}, Vol.~\bibinfo {volume} {2}\
  (\bibinfo  {publisher} {Elsevier Science},\ \bibinfo {address} {Amsterdam},\
  \bibinfo {year} {1996})\ p.\ \bibinfo {pages} {642}\BibitemShut {NoStop}%
\bibitem [{\citenamefont {Zwanzig}(2001)}]{Zwanzig2001}%
  \BibitemOpen
  \bibfield  {author} {\bibinfo {author} {\bibfnamefont {R.}~\bibnamefont
  {Zwanzig}},\ }\href@noop {} {\emph {\bibinfo {title} {Nonequilibrium
  Statistical Mechanics}}},\ \bibinfo {edition} {3rd}\ ed.\ (\bibinfo
  {publisher} {Oxford University Press},\ \bibinfo {address} {New York},\
  \bibinfo {year} {2001})\ p.\ \bibinfo {pages} {240}\BibitemShut {NoStop}%
\bibitem [{\citenamefont {Wittkowski}\ \emph {et~al.}(2012)\citenamefont
  {Wittkowski}, \citenamefont {L{\"o}wen},\ and\ \citenamefont
  {Brand}}]{WittkowskiLB2012}%
  \BibitemOpen
  \bibfield  {author} {\bibinfo {author} {\bibfnamefont {R.}~\bibnamefont
  {Wittkowski}}, \bibinfo {author} {\bibfnamefont {H.}~\bibnamefont
  {L{\"o}wen}}, \ and\ \bibinfo {author} {\bibfnamefont {H.~R.}\ \bibnamefont
  {Brand}},\ }\href@noop {} {\bibfield  {journal} {\bibinfo  {journal} {Journal
  of Chemical Physics}\ }\textbf {\bibinfo {volume} {-- in press}} (\bibinfo
  {year} {2012})}\BibitemShut {NoStop}%
\bibitem [{\citenamefont {Marconi}\ and\ \citenamefont
  {Tarazona}(1999)}]{MarconiT1999}%
  \BibitemOpen
  \bibfield  {author} {\bibinfo {author} {\bibfnamefont {U.~M.~B.}\
  \bibnamefont {Marconi}}\ and\ \bibinfo {author} {\bibfnamefont
  {P.}~\bibnamefont {Tarazona}},\ }\href@noop {} {\bibfield  {journal}
  {\bibinfo  {journal} {Journal of Chemical Physics}\ }\textbf {\bibinfo
  {volume} {110}},\ \bibinfo {pages} {8032} (\bibinfo {year}
  {1999})}\BibitemShut {NoStop}%
\bibitem [{\citenamefont {{Marconi}}\ and\ \citenamefont
  {{Tarazona}}(2000)}]{MarconiT2000}%
  \BibitemOpen
  \bibfield  {author} {\bibinfo {author} {\bibfnamefont {U.~M.~B.}\
  \bibnamefont {{Marconi}}}\ and\ \bibinfo {author} {\bibfnamefont
  {P.}~\bibnamefont {{Tarazona}}},\ }\href@noop {} {\bibfield  {journal}
  {\bibinfo  {journal} {Journal of Physics: Condensed Matter}\ }\textbf
  {\bibinfo {volume} {12}},\ \bibinfo {pages} {413} (\bibinfo {year}
  {2000})}\BibitemShut {NoStop}%
\bibitem [{\citenamefont {Archer}\ and\ \citenamefont
  {Evans}(2004)}]{ArcherE2004}%
  \BibitemOpen
  \bibfield  {author} {\bibinfo {author} {\bibfnamefont {A.~J.}\ \bibnamefont
  {Archer}}\ and\ \bibinfo {author} {\bibfnamefont {R.}~\bibnamefont {Evans}},\
  }\href@noop {} {\bibfield  {journal} {\bibinfo  {journal} {Journal of
  Chemical Physics}\ }\textbf {\bibinfo {volume} {121}},\ \bibinfo {pages}
  {4246} (\bibinfo {year} {2004})}\BibitemShut {NoStop}%
\bibitem [{\citenamefont {Espa{\~n}ol}\ and\ \citenamefont
  {L{\"o}wen}(2009)}]{EspanolL2009}%
  \BibitemOpen
  \bibfield  {author} {\bibinfo {author} {\bibfnamefont {P.}~\bibnamefont
  {Espa{\~n}ol}}\ and\ \bibinfo {author} {\bibfnamefont {H.}~\bibnamefont
  {L{\"o}wen}},\ }\href@noop {} {\bibfield  {journal} {\bibinfo  {journal}
  {Journal of Chemical Physics}\ }\textbf {\bibinfo {volume} {131}},\ \bibinfo
  {pages} {244101} (\bibinfo {year} {2009})}\BibitemShut {NoStop}%
\bibitem [{\citenamefont {Wittkowski}\ and\ \citenamefont
  {L{\"o}wen}(2011)}]{WittkowskiL2011}%
  \BibitemOpen
  \bibfield  {author} {\bibinfo {author} {\bibfnamefont {R.}~\bibnamefont
  {Wittkowski}}\ and\ \bibinfo {author} {\bibfnamefont {H.}~\bibnamefont
  {L{\"o}wen}},\ }\href@noop {} {\bibfield  {journal} {\bibinfo  {journal}
  {Molecular Physics}\ }\textbf {\bibinfo {volume} {109}},\ \bibinfo {pages}
  {2935} (\bibinfo {year} {2011})}\BibitemShut {NoStop}%
\bibitem [{\citenamefont {Wittkowski}\ \emph {et~al.}(2010)\citenamefont
  {Wittkowski}, \citenamefont {L{\"o}wen},\ and\ \citenamefont
  {Brand}}]{WittkowskiLB2010}%
  \BibitemOpen
  \bibfield  {author} {\bibinfo {author} {\bibfnamefont {R.}~\bibnamefont
  {Wittkowski}}, \bibinfo {author} {\bibfnamefont {H.}~\bibnamefont
  {L{\"o}wen}}, \ and\ \bibinfo {author} {\bibfnamefont {H.~R.}\ \bibnamefont
  {Brand}},\ }\href@noop {} {\bibfield  {journal} {\bibinfo  {journal}
  {Physical Review E}\ }\textbf {\bibinfo {volume} {82}},\ \bibinfo {pages}
  {031708} (\bibinfo {year} {2010})}\BibitemShut {NoStop}%
\bibitem [{\citenamefont {Wittkowski}\ \emph
  {et~al.}(2011{\natexlab{a}})\citenamefont {Wittkowski}, \citenamefont
  {L{\"o}wen},\ and\ \citenamefont {Brand}}]{WittkowskiLB2011}%
  \BibitemOpen
  \bibfield  {author} {\bibinfo {author} {\bibfnamefont {R.}~\bibnamefont
  {Wittkowski}}, \bibinfo {author} {\bibfnamefont {H.}~\bibnamefont
  {L{\"o}wen}}, \ and\ \bibinfo {author} {\bibfnamefont {H.~R.}\ \bibnamefont
  {Brand}},\ }\href@noop {} {\bibfield  {journal} {\bibinfo  {journal}
  {Physical Review E}\ }\textbf {\bibinfo {volume} {83}},\ \bibinfo {pages}
  {061706} (\bibinfo {year} {2011}{\natexlab{a}})}\BibitemShut {NoStop}%
\bibitem [{\citenamefont {Wittkowski}\ \emph
  {et~al.}(2011{\natexlab{b}})\citenamefont {Wittkowski}, \citenamefont
  {L{\"o}wen},\ and\ \citenamefont {Brand}}]{WittkowskiLB2011b}%
  \BibitemOpen
  \bibfield  {author} {\bibinfo {author} {\bibfnamefont {R.}~\bibnamefont
  {Wittkowski}}, \bibinfo {author} {\bibfnamefont {H.}~\bibnamefont
  {L{\"o}wen}}, \ and\ \bibinfo {author} {\bibfnamefont {H.~R.}\ \bibnamefont
  {Brand}},\ }\href@noop {} {\bibfield  {journal} {\bibinfo  {journal}
  {Physical Review E}\ }\textbf {\bibinfo {volume} {84}},\ \bibinfo {pages}
  {041708} (\bibinfo {year} {2011}{\natexlab{b}})}\BibitemShut {NoStop}%
\bibitem [{\citenamefont {Emmerich}\ \emph {et~al.}(2012)\citenamefont
  {Emmerich}, \citenamefont {L{\"o}wen}, \citenamefont {Wittkowski},
  \citenamefont {Gruhn}, \citenamefont {T{\'o}th}, \citenamefont {Tegze},\ and\
  \citenamefont {Gr{\'a}n{\'a}sy}}]{EmmerichEtAl2012}%
  \BibitemOpen
  \bibfield  {author} {\bibinfo {author} {\bibfnamefont {H.}~\bibnamefont
  {Emmerich}}, \bibinfo {author} {\bibfnamefont {H.}~\bibnamefont {L{\"o}wen}},
  \bibinfo {author} {\bibfnamefont {R.}~\bibnamefont {Wittkowski}}, \bibinfo
  {author} {\bibfnamefont {T.}~\bibnamefont {Gruhn}}, \bibinfo {author}
  {\bibfnamefont {G.~I.}\ \bibnamefont {T{\'o}th}}, \bibinfo {author}
  {\bibfnamefont {G.}~\bibnamefont {Tegze}}, \ and\ \bibinfo {author}
  {\bibfnamefont {L.}~\bibnamefont {Gr{\'a}n{\'a}sy}},\ }\href@noop {}
  {\bibfield  {journal} {\bibinfo  {journal} {Advances in Physics}\ }\textbf
  {\bibinfo {volume} {61}},\ \bibinfo {pages} {665} (\bibinfo {year}
  {2012})}\BibitemShut {NoStop}%
\bibitem [{\citenamefont {Pleiner}\ and\ \citenamefont
  {Brand}(1996)}]{PleinerB1996}%
  \BibitemOpen
  \bibfield  {author} {\bibinfo {author} {\bibfnamefont {H.}~\bibnamefont
  {Pleiner}}\ and\ \bibinfo {author} {\bibfnamefont {H.~R.}\ \bibnamefont
  {Brand}},\ }in\ \href@noop {} {\emph {\bibinfo {booktitle} {Pattern Formation
  in Liquid Crystals}}},\ \bibinfo {series} {Partially Ordered Systems},
  Vol.~\bibinfo {volume} {9},\ \bibinfo {editor} {edited by\ \bibinfo {editor}
  {\bibfnamefont {A.}~\bibnamefont {Buka}}\ and\ \bibinfo {editor}
  {\bibfnamefont {L.}~\bibnamefont {Kramer}}}\ (\bibinfo  {publisher}
  {Springer},\ \bibinfo {address} {New York},\ \bibinfo {year} {1996})\
  \bibinfo {edition} {1st}\ ed.,\ \bibinfo {type} {Chapter}~\bibinfo {chapter}
  {2}, pp.\ \bibinfo {pages} {15--67}\BibitemShut {NoStop}%
\bibitem [{\citenamefont {Graham}\ and\ \citenamefont
  {Haken}(1971{\natexlab{a}})}]{GrahamH1971a}%
  \BibitemOpen
  \bibfield  {author} {\bibinfo {author} {\bibfnamefont {R.}~\bibnamefont
  {Graham}}\ and\ \bibinfo {author} {\bibfnamefont {H.}~\bibnamefont {Haken}},\
  }\href@noop {} {\bibfield  {journal} {\bibinfo  {journal} {Zeitschrift
  f{\"u}r Physik}\ }\textbf {\bibinfo {volume} {245}},\ \bibinfo {pages} {141}
  (\bibinfo {year} {1971}{\natexlab{a}})}\BibitemShut {NoStop}%
\bibitem [{\citenamefont {Graham}\ and\ \citenamefont
  {Haken}(1971{\natexlab{b}})}]{GrahamH1971b}%
  \BibitemOpen
  \bibfield  {author} {\bibinfo {author} {\bibfnamefont {R.}~\bibnamefont
  {Graham}}\ and\ \bibinfo {author} {\bibfnamefont {H.}~\bibnamefont {Haken}},\
  }\href@noop {} {\bibfield  {journal} {\bibinfo  {journal} {Zeitschrift
  f{\"u}r Physik}\ }\textbf {\bibinfo {volume} {243}},\ \bibinfo {pages} {289}
  (\bibinfo {year} {1971}{\natexlab{b}})}\BibitemShut {NoStop}%
\bibitem [{\citenamefont {Risken}(1972)}]{Risken1972}%
  \BibitemOpen
  \bibfield  {author} {\bibinfo {author} {\bibfnamefont {H.}~\bibnamefont
  {Risken}},\ }\href@noop {} {\bibfield  {journal} {\bibinfo  {journal}
  {Zeitschrift f{\"u}r Physik}\ }\textbf {\bibinfo {volume} {251}},\ \bibinfo
  {pages} {231} (\bibinfo {year} {1972})}\BibitemShut {NoStop}%
\bibitem [{\citenamefont {Risken}(1996)}]{Risken1996}%
  \BibitemOpen
  \bibfield  {author} {\bibinfo {author} {\bibfnamefont {H.}~\bibnamefont
  {Risken}},\ }\href@noop {} {\emph {\bibinfo {title} {The Fokker-Planck
  Equation: Methods of Solution and Applications}}},\ \bibinfo {edition} {3rd}\
  ed.,\ \bibinfo {series} {Springer Series in Synergetics}, Vol.~\bibinfo
  {volume} {18}\ (\bibinfo  {publisher} {Springer},\ \bibinfo {address}
  {Berlin},\ \bibinfo {year} {1996})\ p.\ \bibinfo {pages} {474}\BibitemShut
  {NoStop}%
\bibitem [{\citenamefont {{Kadanoff}}\ and\ \citenamefont
  {{Martin}}(1963)}]{KadanoffM1963}%
  \BibitemOpen
  \bibfield  {author} {\bibinfo {author} {\bibfnamefont {L.~P.}\ \bibnamefont
  {{Kadanoff}}}\ and\ \bibinfo {author} {\bibfnamefont {P.~C.}\ \bibnamefont
  {{Martin}}},\ }\href@noop {} {\bibfield  {journal} {\bibinfo  {journal}
  {Annals of Physics}\ }\textbf {\bibinfo {volume} {24}},\ \bibinfo {pages}
  {419} (\bibinfo {year} {1963})}\BibitemShut {NoStop}%
\bibitem [{\citenamefont {{Hohenberg}}\ and\ \citenamefont
  {{Martin}}(1965)}]{HohenbergM1965}%
  \BibitemOpen
  \bibfield  {author} {\bibinfo {author} {\bibfnamefont {P.~C.}\ \bibnamefont
  {{Hohenberg}}}\ and\ \bibinfo {author} {\bibfnamefont {P.~C.}\ \bibnamefont
  {{Martin}}},\ }\href@noop {} {\bibfield  {journal} {\bibinfo  {journal}
  {Annals of Physics}\ }\textbf {\bibinfo {volume} {34}},\ \bibinfo {pages}
  {291} (\bibinfo {year} {1965})}\BibitemShut {NoStop}%
\bibitem [{\citenamefont {{Brand}}\ \emph {et~al.}(1979)\citenamefont
  {{Brand}}, \citenamefont {{D{\"o}rfle}},\ and\ \citenamefont
  {{Graham}}}]{BrandDG1979}%
  \BibitemOpen
  \bibfield  {author} {\bibinfo {author} {\bibfnamefont {H.}~\bibnamefont
  {{Brand}}}, \bibinfo {author} {\bibfnamefont {M.}~\bibnamefont
  {{D{\"o}rfle}}}, \ and\ \bibinfo {author} {\bibfnamefont {R.}~\bibnamefont
  {{Graham}}},\ }\href@noop {} {\bibfield  {journal} {\bibinfo  {journal}
  {Annals of Physics}\ }\textbf {\bibinfo {volume} {119}},\ \bibinfo {pages}
  {434} (\bibinfo {year} {1979})}\BibitemShut {NoStop}%
\bibitem [{\citenamefont {{Callen}}\ and\ \citenamefont
  {{Welton}}(1951)}]{CallenW1951}%
  \BibitemOpen
  \bibfield  {author} {\bibinfo {author} {\bibfnamefont {H.~B.}\ \bibnamefont
  {{Callen}}}\ and\ \bibinfo {author} {\bibfnamefont {T.~A.}\ \bibnamefont
  {{Welton}}},\ }\href@noop {} {\bibfield  {journal} {\bibinfo  {journal}
  {Physical Review}\ }\textbf {\bibinfo {volume} {83}},\ \bibinfo {pages} {34}
  (\bibinfo {year} {1951})}\BibitemShut {NoStop}%
\bibitem [{\citenamefont {Chaikin}\ and\ \citenamefont
  {Lubensky}(1995)}]{ChaikinL1995}%
  \BibitemOpen
  \bibfield  {author} {\bibinfo {author} {\bibfnamefont {P.~M.}\ \bibnamefont
  {Chaikin}}\ and\ \bibinfo {author} {\bibfnamefont {T.~C.}\ \bibnamefont
  {Lubensky}},\ }\href@noop {} {\emph {\bibinfo {title} {Principles of
  Condensed Matter Physics}}},\ \bibinfo {edition} {1st}\ ed.\ (\bibinfo
  {publisher} {Cambridge University Press},\ \bibinfo {address} {Cambridge},\
  \bibinfo {year} {1995})\ p.\ \bibinfo {pages} {699}\BibitemShut {NoStop}%
\bibitem [{\citenamefont {Elder}\ \emph {et~al.}(2007)\citenamefont {Elder},
  \citenamefont {Provatas}, \citenamefont {Berry}, \citenamefont {Stefanovic},\
  and\ \citenamefont {Grant}}]{ElderPBSG2007}%
  \BibitemOpen
  \bibfield  {author} {\bibinfo {author} {\bibfnamefont {K.~R.}\ \bibnamefont
  {Elder}}, \bibinfo {author} {\bibfnamefont {N.}~\bibnamefont {Provatas}},
  \bibinfo {author} {\bibfnamefont {J.}~\bibnamefont {Berry}}, \bibinfo
  {author} {\bibfnamefont {P.}~\bibnamefont {Stefanovic}}, \ and\ \bibinfo
  {author} {\bibfnamefont {M.}~\bibnamefont {Grant}},\ }\href@noop {}
  {\bibfield  {journal} {\bibinfo  {journal} {Physical Review B}\ }\textbf
  {\bibinfo {volume} {75}},\ \bibinfo {pages} {064107} (\bibinfo {year}
  {2007})}\BibitemShut {NoStop}%
\bibitem [{\citenamefont {van Teeffelen}\ \emph {et~al.}(2009)\citenamefont
  {van Teeffelen}, \citenamefont {Backofen}, \citenamefont {Voigt},\ and\
  \citenamefont {L{\"o}wen}}]{vanTeeffelenBVL2009}%
  \BibitemOpen
  \bibfield  {author} {\bibinfo {author} {\bibfnamefont {S.}~\bibnamefont {van
  Teeffelen}}, \bibinfo {author} {\bibfnamefont {R.}~\bibnamefont {Backofen}},
  \bibinfo {author} {\bibfnamefont {A.}~\bibnamefont {Voigt}}, \ and\ \bibinfo
  {author} {\bibfnamefont {H.}~\bibnamefont {L{\"o}wen}},\ }\href@noop {}
  {\bibfield  {journal} {\bibinfo  {journal} {Physical Review E}\ }\textbf
  {\bibinfo {volume} {79}},\ \bibinfo {pages} {051404} (\bibinfo {year}
  {2009})}\BibitemShut {NoStop}%
\bibitem [{\citenamefont {L{\"o}wen}(2010)}]{Loewen2010}%
  \BibitemOpen
  \bibfield  {author} {\bibinfo {author} {\bibfnamefont {H.}~\bibnamefont
  {L{\"o}wen}},\ }\href@noop {} {\bibfield  {journal} {\bibinfo  {journal}
  {Journal of Physics: Condensed Matter}\ }\textbf {\bibinfo {volume} {22}},\
  \bibinfo {pages} {364105} (\bibinfo {year} {2010})}\BibitemShut {NoStop}%
\bibitem [{\citenamefont {Oxtoby}(1991)}]{Oxtoby1991}%
  \BibitemOpen
  \bibfield  {author} {\bibinfo {author} {\bibfnamefont {D.}~\bibnamefont
  {Oxtoby}},\ }in\ \href@noop {} {\emph {\bibinfo {booktitle} {Liquids,
  Freezing and Glass Transition}}},\ \bibinfo {series} {Proceedings of the
  {L}es {H}ouches {S}ummer {S}chool, {C}ourse {LI}, 3-28 {J}uly 1989},
  Vol.~\bibinfo {volume} {1},\ \bibinfo {editor} {edited by\ \bibinfo {editor}
  {\bibfnamefont {J.-P.}\ \bibnamefont {Hansen}}, \bibinfo {editor}
  {\bibfnamefont {D.}~\bibnamefont {Levesque}}, \ and\ \bibinfo {editor}
  {\bibfnamefont {J.}~\bibnamefont {{Zinn-Justin}}}},\ \bibinfo {organization}
  {{USMG}, {NATO} {A}dvanced {S}tudy {I}nstitute}\ (\bibinfo  {publisher}
  {North {H}olland, {E}lsevier {S}cience {P}ublishers {B}. {V}.},\ \bibinfo
  {address} {Amsterdam},\ \bibinfo {year} {1991})\ pp.\ \bibinfo {pages}
  {145--192}\BibitemShut {NoStop}%
\bibitem [{\citenamefont {Singh}(1991)}]{Singh1991}%
  \BibitemOpen
  \bibfield  {author} {\bibinfo {author} {\bibfnamefont {Y.}~\bibnamefont
  {Singh}},\ }\href@noop {} {\bibfield  {journal} {\bibinfo  {journal} {Physics
  Reports}\ }\textbf {\bibinfo {volume} {207}},\ \bibinfo {pages} {351}
  (\bibinfo {year} {1991})}\BibitemShut {NoStop}%
\bibitem [{\citenamefont {L{\"o}wen}(1994)}]{Loewen1994a}%
  \BibitemOpen
  \bibfield  {author} {\bibinfo {author} {\bibfnamefont {H.}~\bibnamefont
  {L{\"o}wen}},\ }\href@noop {} {\bibfield  {journal} {\bibinfo  {journal}
  {Physics Reports}\ }\textbf {\bibinfo {volume} {237}},\ \bibinfo {pages}
  {249} (\bibinfo {year} {1994})}\BibitemShut {NoStop}%
\bibitem [{\citenamefont {Khalatnikov}(1989)}]{Khalatnikov1989}%
  \BibitemOpen
  \bibfield  {author} {\bibinfo {author} {\bibfnamefont {I.~M.}\ \bibnamefont
  {Khalatnikov}},\ }\href@noop {} {\emph {\bibinfo {title} {An Introduction to
  the Theory of Superfluidity}}},\ \bibinfo {edition} {2nd}\ ed.,\ \bibinfo
  {series} {Frontiers in Physics}, Vol.~\bibinfo {volume} {23}\ (\bibinfo
  {publisher} {Addison Wesley},\ \bibinfo {address} {Redwood City},\ \bibinfo
  {year} {1989})\ p.\ \bibinfo {pages} {206}\BibitemShut {NoStop}%
\bibitem [{\citenamefont {Graham}\ and\ \citenamefont
  {Pleiner}(1975)}]{GrahamP1975}%
  \BibitemOpen
  \bibfield  {author} {\bibinfo {author} {\bibfnamefont {R.}~\bibnamefont
  {Graham}}\ and\ \bibinfo {author} {\bibfnamefont {H.}~\bibnamefont
  {Pleiner}},\ }\href@noop {} {\bibfield  {journal} {\bibinfo  {journal}
  {Physical Review Letters}\ }\textbf {\bibinfo {volume} {34}},\ \bibinfo
  {pages} {792} (\bibinfo {year} {1975})}\BibitemShut {NoStop}%
\bibitem [{\citenamefont {Brand}\ \emph {et~al.}(1990)\citenamefont {Brand},
  \citenamefont {Pleiner},\ and\ \citenamefont {Renz}}]{BrandPR1990}%
  \BibitemOpen
  \bibfield  {author} {\bibinfo {author} {\bibfnamefont {H.~R.}\ \bibnamefont
  {Brand}}, \bibinfo {author} {\bibfnamefont {H.}~\bibnamefont {Pleiner}}, \
  and\ \bibinfo {author} {\bibfnamefont {W.}~\bibnamefont {Renz}},\ }\href@noop
  {} {\bibfield  {journal} {\bibinfo  {journal} {Journal de Physique}\ }\textbf
  {\bibinfo {volume} {51}},\ \bibinfo {pages} {1065} (\bibinfo {year}
  {1990})}\BibitemShut {NoStop}%
\bibitem [{\citenamefont {{Temmen}}\ \emph {et~al.}(2000)\citenamefont
  {{Temmen}}, \citenamefont {{Pleiner}}, \citenamefont {{Liu}},\ and\
  \citenamefont {{Brand}}}]{TemmenPLB2000}%
  \BibitemOpen
  \bibfield  {author} {\bibinfo {author} {\bibfnamefont {H.}~\bibnamefont
  {{Temmen}}}, \bibinfo {author} {\bibfnamefont {H.}~\bibnamefont {{Pleiner}}},
  \bibinfo {author} {\bibfnamefont {M.}~\bibnamefont {{Liu}}}, \ and\ \bibinfo
  {author} {\bibfnamefont {H.~R.}\ \bibnamefont {{Brand}}},\ }\href@noop {}
  {\bibfield  {journal} {\bibinfo  {journal} {Physical Review Letters}\
  }\textbf {\bibinfo {volume} {84}},\ \bibinfo {pages} {3228} (\bibinfo {year}
  {2000})}\BibitemShut {NoStop}%
\bibitem [{\citenamefont {{Temmen}}\ \emph {et~al.}(2001)\citenamefont
  {{Temmen}}, \citenamefont {{Pleiner}}, \citenamefont {{Liu}},\ and\
  \citenamefont {{Brand}}}]{TemmenPLB2001}%
  \BibitemOpen
  \bibfield  {author} {\bibinfo {author} {\bibfnamefont {H.}~\bibnamefont
  {{Temmen}}}, \bibinfo {author} {\bibfnamefont {H.}~\bibnamefont {{Pleiner}}},
  \bibinfo {author} {\bibfnamefont {M.}~\bibnamefont {{Liu}}}, \ and\ \bibinfo
  {author} {\bibfnamefont {H.~R.}\ \bibnamefont {{Brand}}},\ }\href@noop {}
  {\bibfield  {journal} {\bibinfo  {journal} {Physical Review Letters}\
  }\textbf {\bibinfo {volume} {86}},\ \bibinfo {pages} {745} (\bibinfo {year}
  {2001})}\BibitemShut {NoStop}%
\bibitem [{\citenamefont {Pleiner}\ \emph {et~al.}(2000)\citenamefont
  {Pleiner}, \citenamefont {Liu},\ and\ \citenamefont {Brand}}]{PleinerLB2000}%
  \BibitemOpen
  \bibfield  {author} {\bibinfo {author} {\bibfnamefont {H.}~\bibnamefont
  {Pleiner}}, \bibinfo {author} {\bibfnamefont {M.}~\bibnamefont {Liu}}, \ and\
  \bibinfo {author} {\bibfnamefont {H.~R.}\ \bibnamefont {Brand}},\ }\href@noop
  {} {\bibfield  {journal} {\bibinfo  {journal} {Rheologica Acta}\ }\textbf
  {\bibinfo {volume} {39}},\ \bibinfo {pages} {560} (\bibinfo {year}
  {2000})}\BibitemShut {NoStop}%
\bibitem [{\citenamefont {Pleiner}\ \emph {et~al.}(2002)\citenamefont
  {Pleiner}, \citenamefont {Liu},\ and\ \citenamefont {Brand}}]{PleinerLB2002}%
  \BibitemOpen
  \bibfield  {author} {\bibinfo {author} {\bibfnamefont {H.}~\bibnamefont
  {Pleiner}}, \bibinfo {author} {\bibfnamefont {M.}~\bibnamefont {Liu}}, \ and\
  \bibinfo {author} {\bibfnamefont {H.~R.}\ \bibnamefont {Brand}},\ }\href@noop
  {} {\bibfield  {journal} {\bibinfo  {journal} {Rheologica Acta}\ }\textbf
  {\bibinfo {volume} {41}},\ \bibinfo {pages} {375} (\bibinfo {year}
  {2002})}\BibitemShut {NoStop}%
\bibitem [{\citenamefont {Pleiner}\ \emph {et~al.}(2004)\citenamefont
  {Pleiner}, \citenamefont {Liu},\ and\ \citenamefont {Brand}}]{PleinerLB2004}%
  \BibitemOpen
  \bibfield  {author} {\bibinfo {author} {\bibfnamefont {H.}~\bibnamefont
  {Pleiner}}, \bibinfo {author} {\bibfnamefont {M.}~\bibnamefont {Liu}}, \ and\
  \bibinfo {author} {\bibfnamefont {H.~R.}\ \bibnamefont {Brand}},\ }\href@noop
  {} {\bibfield  {journal} {\bibinfo  {journal} {Rheologica Acta}\ }\textbf
  {\bibinfo {volume} {43}},\ \bibinfo {pages} {502} (\bibinfo {year}
  {2004})}\BibitemShut {NoStop}%
\bibitem [{Esp()}]{Espanol}%
  \BibitemOpen
  \href@noop {} {}\bibinfo {howpublished} {P. Espa\~{n}ol, private
  communication}\BibitemShut {NoStop}%
\bibitem [{\citenamefont {{Schmidt}}(2011)}]{Schmidt2011}%
  \BibitemOpen
  \bibfield  {author} {\bibinfo {author} {\bibfnamefont {M.}~\bibnamefont
  {{Schmidt}}},\ }\href@noop {} {\bibfield  {journal} {\bibinfo  {journal}
  {Physical Review E}\ }\textbf {\bibinfo {volume} {84}},\ \bibinfo {pages}
  {051203} (\bibinfo {year} {2011})}\BibitemShut {NoStop}%
\end{thebibliography}%
\end{document}